\documentclass[manuscript]{acmart}

\AtBeginDocument{%
  \providecommand\BibTeX{{%
    \normalfont B\kern-0.5em{\scshape i\kern-0.25em b}\kern-0.8em\TeX}}}

\setcopyright{acmcopyright}
\copyrightyear{2023}
\acmYear{2023}
\acmDOI{XXXXXXX.XXXXXXX}

\acmConference[CHI'25]{The ACM CHI conference on Human Factors in Computing Systems}{April 26--May 1st,
  2024}{Yokohama, Japan}
\acmPrice{15.00}
\acmISBN{978-1-4503-XXXX-X/18/06}

\usepackage{hyperref}
\usepackage{graphicx}
\usepackage{subcaption}
\usepackage{pdfpages}

\usepackage{lineno}
\usepackage{enumitem}
\usepackage{booktabs}
\usepackage{caption}
\usepackage{array}
\usepackage{enumitem}
\usepackage{multirow}

\begin{document}
\nolinenumbers

\title[Promptimizer]{Promptimizer: User-Led Prompt Optimization for Personal Content Classification}

\author{Leijie Wang}
\email{leijiew@cs.washington.edu}
\affiliation{%
  \institution{University of Washington}
  \city{Seattle}
  \country{United States}
}

\author{Kathryn Yurechko}
\email{kathryn.yurechko@oii.ox.ac.uk}
\affiliation{%
  \institution{University of Oxford}
  \city{Oxford}
  \country{United Kingdom}
}

\author{Amy X. Zhang}
\email{axz@cs.uw.edu}
\affiliation{%
  \institution{University of Washington}
  \city{Seattle}
  \country{United States}
}

\newcommand\deleted[1]{}
\newcommand\leijie[1]{#1}
\newcommand{\System}{Promptimizer}
\newcommand{\YouTubeSystem}{Puffin}
\newcommand{\Username}{Sarah}
\renewcommand{\shortauthors}{Wang Leijie et al.}


\begin{abstract}
While LLMs now enable users to create content classifiers easily through natural language, automatic prompt optimization techniques are often necessary to create performant classifiers.
However, such techniques can fail to consider how social media users want to evolve their filters over the course of usage, including desiring to steer them in different ways during initialization and iteration.
We introduce a user-centered prompt optimization technique, \System{}, that maintains high performance and ease-of-use but additionally (1) allows for user input into the optimization process and (2) produces final prompts that are interpretable.
A lab experiment (n=16) found that users significantly preferred \System{}’s human-in-the-loop optimization over a fully automatic approach.
We further implement \System{} into \textsc{\YouTubeSystem{}}, a tool to support YouTube content creators in creating and maintaining personal classifiers to manage their comments. 
Over a 3-week deployment with 10 creators, participants successfully created diverse filters to better understand their audiences and protect their communities.
\end{abstract}



\maketitle

\begin{figure*}[h!]
    \centering
    \includegraphics[width=\textwidth]{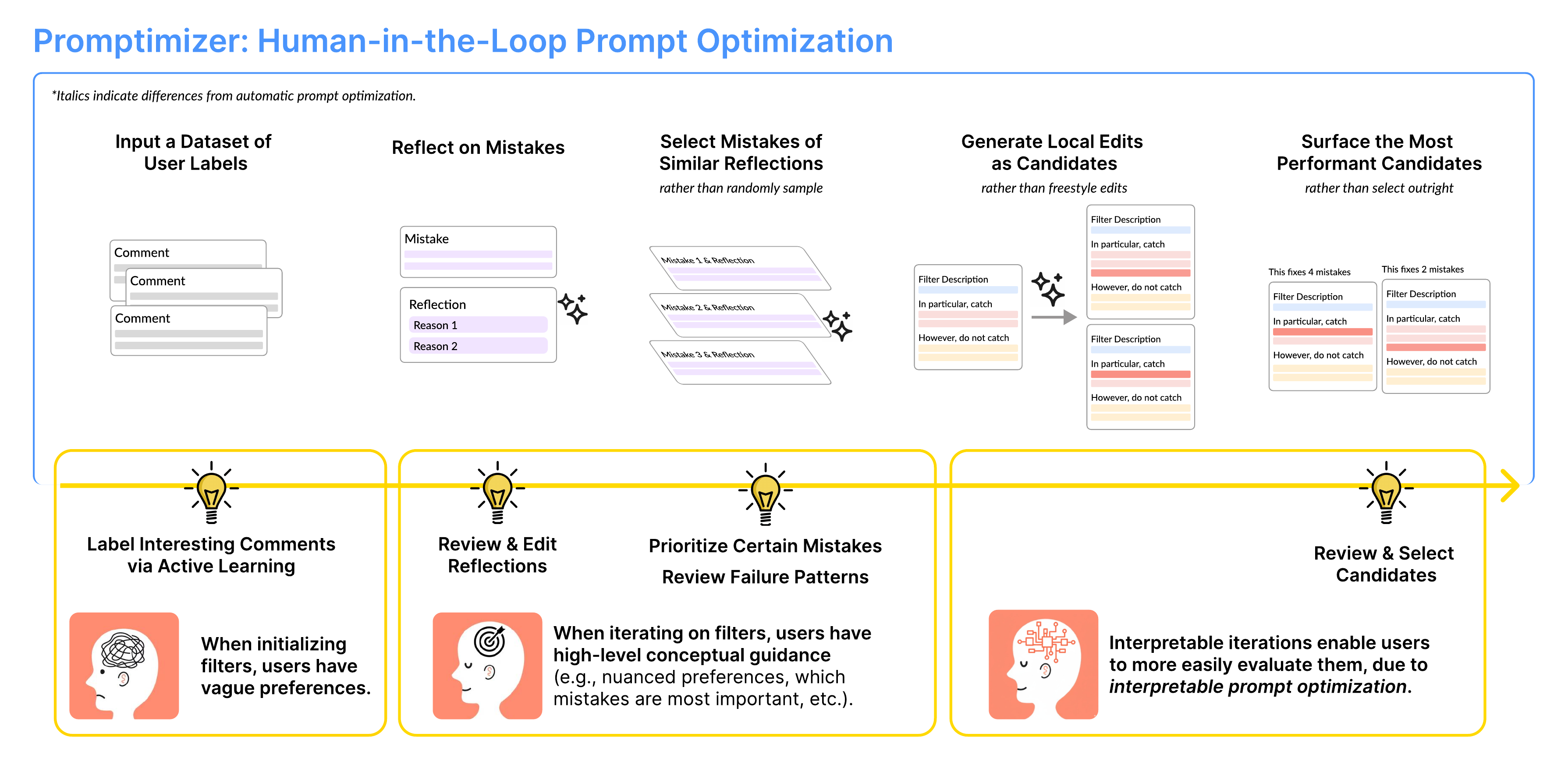}
    \caption{\textbf{Illustration of \System{}: Human-in-the-Loop Prompt Optimization}. 
    In our design of \System{}, we identify opportunities within the prompt optimization workflow where users can communicate their preferences in diverse and evolving ways. After collecting such varied user feedback, \System{} then generates candidate prompts that are performant, generalizable, and interpretable for user review.}
    \hfill
    \label{workflow}
\end{figure*}

\section{Introduction}

Internet users often rely on centralized automated filters to manage a flood of digital content, ranging from moderation algorithms for harmful social-media content~\cite{YoutubeAlgorithm, FacebookAlgorithm} and spam filters for email~\cite{GmailSpamFilter} to curation algorithms for content discovery~\cite{eksombatchai2018pixie}. 
While effective at scale, these one-size-fits-all approaches overlook the diverse preferences of individual users~\cite{neves2018customization,kumar2021designing}. 
To address this gap, researchers have explored tools that support users in creating personalized content classifiers for their own online environments~\cite{jhaver2022designing, he2023cura, park2019opportunities,jhaver2019human}.
These range from rule-based systems where users provide phrases or even regular expressions~\cite{jhaver2022designing,song2023modsandbox, chandrasekharan2019crossmod}, to supervised machine learning models where users provide training data~\cite{smith2020no}, to more recently, large language models (LLMs) where users can prove natural language input~\cite{wang2025end}.


However, existing approaches do not take into account the context of how end users understand and communicate their complex and evolving preferences for content filters.
When users first create a filter, they typically have a vague intuition about what content they want to see or avoid~\cite{kulesza2014structured, chen2021goldilocks, chang2017revolt}. 
For that reason, they may struggle to come up with high-level descriptions or concrete rules, as is typically expected in LLM-based or rule-based systems, though can often make intuitive decisions about content examples~\cite{wang2025end}.
Later, as users see their filter in action, they develop a more targeted understanding of how to refine it~\cite{lee2024clarify, daly2021user, piorkowski2023aimee}.
For instance, social media users discover filtering mistakes as they browse their feeds regularly~\cite{shen2021everyday}.
As they encounter mistakes, users may have specific ideas about what mistakes to focus on correcting or a preference for greater precision or recall~\cite{wang2025end}. 
Besides mistakes, user preferences may also evolve over time, necessitating continuous iterations~\cite{feng2024mapping}.
During this iteration process, users no longer want to label many examples tediously, as would be expected in interactive machine learning systems~\cite{smith2020no}, but prefer to articulate corrections at a conceptual level~\cite{ratner2017snorkel}.

Given this context, we argue that the best solution for users would need to incorporate the ability for \textit{flexible forms of user input}, to cover how users refine their preferences over time, as well as \textit{interpretability} so that users can grasp how their filter makes decisions to be able to diagnose and target corrections~\cite{kulesza2015principles, piorkowski2023aimee, amershi2014power}.
Of the various options, LLMs seem the most suited for taking a range of inputs and also being able to make targeted edits to prompts for iterative refinement~\cite{kumar2023watch, OpenAI2024GPT4}.
But while it is trivial for users to create a decently performing classifier, more complex prompt engineering is needed to capture nuanced preferences~\cite{zamfirescu2023johnny}.
For instance, providing examples often helps improve accuracy but the selection and ordering of these examples can significantly influence performance~\cite{rubin2021learning, zhang2022active, lu2021fantastically}.
Additionally, users struggle to refine LLM prompts at a fine-grained level once created~\cite{zamfirescu2023herding}.

To address these challenges, researchers have proposed automatic prompt optimization (APO) techniques that generate performant prompts given small sets of user labels~\cite{deng2022rlprompt, zhou2022large, prasad2022grips, pryzant2023automatic}.
In each turn, APO algorithms randomly sample a few mistakes, then ask an LLM to modify prompts in a freestyle fashion to fix the mistakes.
While these techniques could be useful to create a new filter from scratch, they do not provide any way for users to give input during iteration beyond adding more labels.
Moreover, their focus on performance on narrowly defined datasets produces prompts that are difficult for users to interpret~\cite{deng2022rlprompt, shin2020autoprompt, shi2022toward} and might even overfit~\cite{li2023robust}.

In this work, we propose \System{}, a human-in-the-loop (HiTL) prompt optimization workflow that supports users in initializing and iterating on content filters.
In our design of \System{}, we seek to enable users to express their complex and evolving preferences through diverse forms of input, without needing to manage the low-level mechanics of prompt optimization. We achieve this via the following two changes compared to fully automated approaches: 

\begin{itemize}
    \item \textbf{User-Led Steering}. During initialization, users provide an initial description and label a small set of examples surfaced through active learning---specifically, examples where LLMs have inconsistent predictions, reflecting potential ambiguities. 
    As users identify mistakes during casual use, \System{} summarizes common failure patterns and asks for user review; users can also prioritize certain mistakes and clarify their preferences.
    
    \item \textbf{Interpretable Prompt Optimization}. After collecting user feedback, \System{} generates candidate prompts that are performant, generalizable, and interpretable for user review.
    To promote interpretability, we analyze which prompt rubric relates most directly to user feedback and constrain LLMs to making generalizable iterations to those rubrics, unlike existing APO approaches that allow unconstrained edits. 
    We then identify the most performant candidates for user selection by using established searching mechanisms in APO algorithms.
\end{itemize}

To understand how \System{} supports users to author performant and interpretable filters, we conducted a within-subjects experiment with 16 social-media users, comparing \System{} against a state-of-the-art APO algorithm. 
We found that participants unanimously preferred \System{} over an APO workflow for filter iteration. 
They also perceived \System{} as producing significantly more interpretable iterations while achieving comparable performance. 
In post-study interviews, participants spoke approvingly about how \System{} gave them more control to pick their iteration strategy, including prioritizing specific mistakes, navigating precision-recall trade-offs, addressing failure patterns, and revising initial labels.


In order to understand how our filter authoring workflow would work in a real social media context, we implemented \System{} into a tool called \textsc{\YouTubeSystem{}} built for YouTube content creators. We chose this context as creators are a class of end users who have a particularly strong need for personalized classifiers over their comments, due to their desire to engage with their audience and moderate out harmful targeted attacks~\cite{jhaver2022designing}.
\textsc{\YouTubeSystem{}} allows a creator to create content filters for comments on their videos, view and monitor comments separated out by their filters, and adjust their filters as they are reviewing comments in the system and finding mistakes. 

We conducted a three-week field study of \textsc{\YouTubeSystem{}} with 10 YouTube content creators with audiences ranging from under 500 to over 1 million subscribers.
We found that \textsc{\YouTubeSystem{}} enabled creators to iteratively build high-performing filters with ease as they naturally reviewed their comments. 
All participants successfully created a diverse range of filters in \textsc{\YouTubeSystem{}} that helped them better understand their audiences and protect their communities over the course of the study. 
Notably, six out of ten participants are continuing to use or have expressed interest in using \textsc{\YouTubeSystem{}} beyond the study.




Looking forward, we envision that the \System{} workflow can generalize beyond content classification to other LLM tasks, particularly open-ended ones where user guidance is crucial~\cite{guo2025pen, wang2025social}.
Prompt optimization also need not be limited to a single user: groups could collectively deliberate on guidance and review generated candidates. Such deliberation supports consensus-building and thus can be especially valuable for aligning LLMs with community values~\cite{feng2024policy, sharma2024experts}.
We see opportunities for \textsc{\YouTubeSystem{}} to support not only creating personal filters but also sharing and evaluating filters within communities that face similar harms or share similar identities. 


\section{Related Work}

\subsection{Personal Content Classification}
Personal content classification is the process by which an individual selects, organizes, and filters their own view of content online~\cite{jhaver2023personalizing, wang2025end}.
Beyond filtering out unwanted content, such as removing harassing or spam emails~\cite{park2019opportunities, mahar2018squadbox} and managing online conversations~\cite{wang2023reporting}, personally classifying content also entails curating interesting content, such as relevant news articles~\cite{volk2024selecting, schmitt2018toomuch} or personalizing content recommendations on social media~\cite{liu2025filtering}. 
Many content classification tools, however, often operate in a centralized manner, enacted top-down by platforms and therefore limiting users' capability to customize content filtering~\cite{neves2018customization}.
In response, researchers have increasingly called for tools that enable users to customize their online content environments, especially on social media~\cite{jhaver2023personalizing, jhaver2023users}. 

When designing personal content classifiers to enact user preferences, it is essential to consider that user preferences are complex and evolving.
Users often begin with abstract ideas about which kinds of content they want to moderate or curate~\cite{kulesza2014structured, chen2021goldilocks, chang2017revolt}. 
Although reviewing content examples can elicit users' intuitive decisions, articulating preferences in holistic descriptions often remains a challenge~\cite{wang2025end}.
However, with time and experiences like discovering filter mistakes through naturally scrolling their social media feeds, users often want to fix the mistakes that they viewed~\cite{shen2021everyday} and tend to cultivate insights into how to refine their filters~\cite{lee2024clarify, daly2021user, piorkowski2023aimee}.
Users' preferences may also change over time, especially amid online information ecosystems that often evolve quickly, necessitating opportunities for iterating on their initial filters~\cite{feng2024mapping}.
In these cases, labeling numerous redundant examples is not an ideal strategy; instead, users prefer to offer higher-level feedback~\cite{ratner2017snorkel}.
Because black-box algorithms conceal how they make decisions, they impede users in making such iterations effectively, as they cannot easily identify problems and make informed improvements~\cite{smith2020no}.
Filters must instead support users in diagnosing misalignments between their complex, evolving preferences and filter decisions by making internal decision-making interpretable~\cite{kulesza2015principles, piorkowski2023aimee, amershi2014power}.

Rather than support complex and evolving user needs, current personal content classifiers only offer a single, fixed mechanism for users to convey their preferences.
Some systems, like Reddit's AutoMod~\cite{jhaver2019human, song2023modsandbox}, allow community moderators to write regular expressions for inappropriate posts~\cite{jhaver2019human}.
Similar keywords-based rule systems are common across social media platforms~\cite{jhaver2022designing}.
Though interpretable, such rule-based classifiers do not align with how users prefer to communicate their preferences.
Users find it difficult to translate abstract initial preferences into high-performing rules and update rules based on specific needs~\cite{jhaver2022designing,song2023modsandbox, chandrasekharan2019crossmod}.
Other systems utilize supervised machine learning (ML), enabling users to provide labeled examples.
They often incorporate pretrained word-embeddings~\cite{devlin2018bert} and active learning~\cite{settles2009active} to learn from user labels more efficiently~\cite{das2013end, kulesza2014structured}.
Numerous others only allow users to adjust a pre-trained filter's sensitivity~\cite{jhaver2023personalizing, Instagram2024, Bodyguard2024}.
These black-box ML classifiers all lack interpretability, hindering users in comprehending filter decisions and iterating on their filters in accordance with their preferences~\cite{piorkowski2023aimee, kulesza2015principles}.

\subsection{User Prompting for Building Content Classifiers}
LLMs have gained immense popularity due to their effectiveness in natural language processing tasks~\cite{OpenAI2024GPT4}. 
Instead of creating a new model for each specific task, users can simply now customize an LLM by providing a prompt with maybe a few examples at runtime, which is known as in-context learning~\cite{brown2020language}.
Despite the apparent simplicity of prompting, users face prompt engineering challenges that impede them in leveraging LLMs to actualize their preferences~\cite{zamfirescu2023johnny}.
Users typically create prompts without a systematic plan, which hinders consistent progress~\cite{zamfirescu2023johnny}. 
Although they can include labeled examples to help articulate their preferences, how they select and order these examples significantly affects performance~\cite{rubin2021learning, zhang2022active, lu2021fantastically}.
After creating filters, users also struggle to refine their filters with granularity~\cite{zamfirescu2023herding}.
For example, when designing prompts to build robust chatbots, users often solve 80\% of the problems but find it very difficult to fix the remaining issues~\cite{zamfirescu2023herding}. 
When using an experimental prompt-driven classifier, users resorted to adding misclassified examples to their prompts or trying to write rule-like prompts in the absence of structured refinement guidance~\cite{wang2025end}.
Ultimately, users become frustrated with LLMs' inconsistent decisions on identical content, struggling to comprehend such mistakes~\cite{lu2021fantastically, errica2024did, zamfirescu2023johnny}.

As a result, a growing body of research investigates how to replace demanding, manual prompt engineering with automatic prompt optimization (APO) techniques, which automatically generate refined prompts based on a dataset of ground-truth user labels~\cite{deng2022rlprompt, zhou2022large, prasad2022grips, pryzant2023automatic}.
One line of APO research explores how to select the best example to add to a prompt, or examine the order in which examples should be added~\cite{lu2021fantastically}.
For instance, active example selection~\cite{zhang2022active} aims to find the most effective demonstrations in prompts via reinforcement learning policies.
However, adding too many examples can conceal high-level patterns and is not sustainable, making it unsuitable for supporting consistent iterations.
Another line of APO involves writing more precise instructions to communicate task requirements.
For simple tasks, prompt optimizations can easily be achieved through keyword and paraphrasing perturbations~\cite{mishra2023promptaid}, whereas more complex tasks require more substantial prompt edits.
In particular, researchers have drawn inspiration from stochastic gradient descent to train ML classifiers.
In each round, researchers randomly sample several mistakes, then ask LLMs to reflect on these mistakes as natural language gradients. LLMs generate a few prompt edits using backpropagation, ultimately evaluating the prompts on the training data to select the best performing candidates~\cite{pryzant2023automatic}.

\subsection{Human-in-the-Loop Prompt Optimization for Personal Content Classification}
\label{related-work-3}
Automatic prompt optimization prioritizes the final performance of resultant prompts and often minimizes any human intervention in the process~\cite{zhou2022large, prasad2022grips, pryzant2023automatic}.
However, APO poses issues for the setting of personal content classification that suggest the need for a more human-in-the-loop prompt optimization approach. 

\subsubsection{Initializing Performant Filters}\label{initializing-performant}
APO widely assumes ground-truth labels that indicate the task requirements that an ideal prompt should incorporate.
Labeling examples can indeed support users in initializing filters: casual social media users, for example, often begin with unclear, intuitive preferences like knowing that they want to avoid ``hateful content''~\cite{jhaver2023personalizing}. 
However, users often encounter much content that is irrelevant to their preferences, requiring a more efficient way to sample examples worth labeling.
For instance, hate speech on Twitter/X at most accounts for less than 2\% of all posts~\cite{tonneau2024hateday}, whereas around 6\% of Reddit comments violated a broad set of platform norms~\cite{park2022measuring}.
Drawing inspirations from active learning, we prioritize and present the most valuable examples~\cite{lewis1995sequential, seung1992query} for user review during filter initialization.

\subsubsection{Iterating with Diverse Approaches}\label{diverse-iterating}
The ways in which users prefer to iterate on their content filters do not align with APO algorithm requests for labeled examples.
Though examples can communicate intuitive preferences initially, users may develop a more clear and targeted understanding of how to refine their filters over time~\cite{shen2021everyday}.
In these cases, an easy iteration for users would allow them to provide high-level conceptual guidance, whereas labeling repetitive examples becomes less straightforward~\cite{wang2025end}.
Some systems, like ConstitutionalMaker~\cite{petridis2024constitutionmaker}, allow 
diverse user inputs, but such inputs are not used for prompt optimization and suffer from conflicting feedback.
Meanwhile, high-level feedback is valuable because users can have a better idea than algorithms about which filter mistakes should be prioritized in iterations, potentially leading to better perceptions of algorithmic performance.
However, APO is not designed to incorporate this kind of nuanced, high-level input.
It solely optimizes on user labels, and when iterating, the only approach is to run APO for more rounds, excluding users from deciding which mistakes are most important or offering additional guidance.

\subsubsection{Promoting Interpretability}\label{interpretable-optimize}
APO involves LLMs articulating why a small number of randomly selected mistakes were misclassified, attempting to fix the mistakes by generating a few prompt candidates in a freestyle fashion~\cite{zhou2022large, prasad2022grips, pryzant2023automatic}.
While this method promotes high performance, its lack of structure in generating prompt edits often leads to less interpretable prompt iterations: they sample a few mistakes from the training examples, then ask an LLM for free-form edits~\cite{zhou2022large, prasad2022grips, pryzant2023automatic}.
Without a holistic understanding of underlying failure patterns, this unguided process tends to produce lengthy, opaque changes~\cite{deng2022rlprompt, shin2020autoprompt, shi2022toward}.
For instance, LLMs might make sweeping edits throughout a prompt to address distinct problems at one time. 
In some cases, APO might even overfit to individual mistakes by turning all mistakes into few-shot demonstrations~\cite{li2023robust}.
Interpretability also builds user trust in personal content filters and enables meaningful user control over information consumption~\cite{lim2009and}. 
For instance, some social media users worry that black-box filters might create echo chambers~\cite{jhaver2023personalizing}, while numerous content creators fear these systems could mistakenly remove legitimate audience comments and damage their reputation~\cite{jhaver2022designing}.

\section{\System{}: Human-in-the-Loop Prompt Optimization}
We introduce \System{}, a novel human-in-the-loop prompt optimization workflow, that supports users in easily initializing and iterating on performant content classifiers. 
We consider initializing and iterating on personal content classifiers as two distinct scenarios, where how users communicate their preferences can vary greatly. Derived from our arguments in the related work section, we present three design goals for \System{}:

\begin{enumerate}
    \item \textbf{Support users in initializing performant content filters easily.}
    \item \textbf{Enable users to iterate on content filters easily and accurately in diverse ways.}
    \item \textbf{Provide interpretable content filters for users.}
\end{enumerate}

As illustrated in Figure \ref{workflow}, \System{} instantiates these design goals through three interconnected algorithmic components. We now describe these components in detail.

\subsection{\textbf{Performant Classification Algorithms}}
Since prompt optimization seeks the most performant prompt for content classification, it depends heavily on the underlying classification algorithm's inherent performance. 
The interpretability of classification algorithms also affects how accurately users can provide guidance and how effectively the optimization algorithm can correct mistakes.

We first consider: \textbf{how should we structure prompts to ensure interpretability and enable targeted user corrections?} 
We structured each user-defined prompt in three parts: an overall description, positive rubrics, and negative rubrics.
These rubrics help decompose nuanced, complex preferences into more interpretable components~\cite{lee2024clarify, toubal2024modeling, louie2024roleplay}.
This structure also enables LLMs to make focused improvements during optimization rather than wholesale prompt rewrites.
We additionally include four few-shot examples, as research has demonstrated performance gains from this approach~\cite{brown2020language}.
We limit examples to four because larger numbers reduce interpretability for users while only marginally improving performance~\cite{zhao2021calibrate}.

Second, we ask: \textbf{how should classification algorithms produce performant and consistent predictions?}
While high-quality predictions can be obtained with more computational resources, we needed to balance prediction quality with cost.
We experimented with different algorithmic structures (e.g., chain of thought, aggregating predictions from individual rubrics) and parameters using the Jigsaw toxicity dataset~\cite{borkan2019nuanced}, selecting the approach with the highest test performance.
Our final algorithm combines user-created prompts with a predefined system prompt and processes comments in batches of five. 
For each batch, we run the prediction process five times with randomized comment orderings, then aggregate results using majority voting.
This approach addresses several key challenges.
Prior work shows that users become frustrated when identical prompts yield inconsistent decisions~\cite{wang2025end}.
Majority voting reduces this randomness and delivers more consistent, interpretable predictions~\cite{wang2022self}.
Additionally, the level of agreement across runs provides a natural confidence measure, as low-confidence predictions often signal ambiguous user preferences and warrant user review~\cite{yue2023large}.
While majority voting requires more computational resources, we mitigate costs through batched prompting, which preserves performance while reducing overhead~\cite{cheng2023batch, lin2023batchprompt}.
Randomizing comment order across runs prevents predictions from being influenced by positional effects or comment context~\cite{lu2021fantastically}.

\subsection{\textbf{Interpretable Prompt Optimization}}
We focus on a simplified prompt optimization setting: \textbf{Given a small set of user labels and an initial prompt, how do we produce a prompt iteration that is both performant and interpretable?}
As we argued in Section \ref{interpretable-optimize}, standard APO approaches tend to generate less interpretable iterations.
In contrast, our algorithm addresses interpretability issues through structured iteration generation.
First, we identify mistakes that the initial prompt made on training examples and elicit LLM reflections on why each mistake occurred. 
We then cluster these reflections by semantic similarity using unsupervised clustering~\cite{khan2014dbscan}.
This ensures that each cluster represents a coherent failure pattern that can be addressed through focused edits. 
Since prompt filters decompose into positive and negative rubrics, we constrain potential iterations to four directions: adding a new positive or negative rubric, or editing an existing positive or negative rubric. 
For each mistake cluster, we ask the LLM to generate targeted iterations within these constraints. 
Finally, we select the top iterations by evaluating their performance on the training set and present them for user review.

\subsection{\textbf{Efficient Collection of Diverse User Guidance}}  


As discussed in Section \ref{initializing-performant}, standard APO methods assume the availability of relevant ground-truth annotations, but content curation involves much content that is irrelevant to users' preferences.
This raises the question: \textbf{How should we collect user labels most efficiently?} 
We draw inspiration from active learning~\cite{lewis1995sequential, seung1992query}, where uncertainty sampling prioritizes the most informative examples for user labeling~\cite{lewis1995sequential, seung1992query}.
Here, uncertainty arises naturally from our classification algorithm: each comment's label is predicted five times, and comments with inconsistent outcomes across rounds reflect classifier uncertainty.
These comments are more likely to expose ambiguities in the current prompt that could benefit most from user feedback.
When too few uncertain cases exist, we prioritize positive predictions over negative ones, since positive examples are rare and therefore more valuable for initialization~\cite{tonneau2024hateday, park2022measuring}.

Additionally, users want to provide high-level conceptual guidance rather than detailed labels during iterations as discussed in Section \ref{diverse-iterating}.
We therefore ask: \textbf{How can we help users easily provide diverse and nuanced guidance?}
Our interpretable prompt optimization workflow supports flexible user involvement at three points. 
First, noting that our optimization algorithm clusters mistakes with similar failure reasons, we summarize each cluster into a failure pattern for user review.
Second, users might want to address a specific mistake that they view as especially important.
Because optimizing on a single mistake risks overfitting, we ask users to clarify their intent in natural language. 
To minimize effort, they edit the LLM’s reflection on the mistake instead of starting from scratch. 
Because generated iterations are constrained to interpretable edits, users can evaluate candidates not only by performance on a small dataset but also by how well the candidates seem to align with their preferences.


\section{Lab Evaluation of \System{}}
We conducted a within-subjects experiment with 16 social-media users to compare \System{} against an automatic prompt optimization baseline, formulating the following research questions:

\begin{itemize}
    \item \textbf{RQ1}: Does \System{} support users to author more interpretable filters?
    
    \item \textbf{RQ2}: Does \System{} support users to author more performant filters?

    \item \textbf{RQ3}: Do users prefer \System{} for authoring their content filters?

\end{itemize}

\subsection{Recruitment and Participants}
We recruited 16 social-media users, labeled E1–E16, by advertising calls for participation through instant-messaging channels to students at six major universities in the U.S. and U.K. The group included nine female and seven male-identifying participants, with most pursuing graduate degrees and a few pursuing bachelor’s degrees. 
Regarding generative AI, two participants reported that they used it every day, 6 every few days, 2 every few weeks, and 6 never or almost never.
Sessions were conducted individually over Zoom and lasted an average of 93 minutes. 
Participants received a \$20 gift card as compensation. 
This study was reviewed by our university IRB and deemed exempt.

\subsection{Study Procedures and Evaluation Metrics}
\subsubsection{\textbf{Experiment Datasets}}
\label{datasets}
Participants created filters on two topics: political comments from \textit{TheYoungTurks} YouTube channel and food-related comments from YouTube's \textit{BestEverFoodReviewShow}. For politics, the target description was about political hate speech: ``\textit{Comments that spread hate or hostility toward people because of who they are, especially in political discussions.}'' For food, it involved cultural food experiences: ``\textit{Comments where people share detailed and rich cultural experience and knowledge related to food, cooking, or dining.}” 
We selected these topics to represent moderation and curation tasks, respectively. 
Both descriptions contain intentionally vague language to allow participants to develop their own nuanced preferences.
Sampling from real comments allowed us to better simulate real-world curation settings.
We crawled all comments from the most recent 20 videos on each channel.
We then used LLMs in a zero-shot setting to classify whether each comment matched the target description.
Although participants' understanding of each topic can differ greatly, we aimed to balance the dataset so that nearly half of the comments would be in-scope as determined by LLM classifiers. 
From this pool, we randomly sampled 200 comments per topic. 
For each participant, we split these into three sets.

\begin{itemize}
    \item A training set of 20 comments for initializing filters. We ensured that it included at least 10 uncertain cases (inconsistent positive and negative predictions) to support our active learning requirement. 

    \item An audit set of 80 comments to simulate daily filter audits for iterative improvement.

    \item A test set of 100 comments that were later fully labeled by participants to evaluate their final classifiers.
\end{itemize}

\subsubsection{\textbf{Experiment Conditions}}

We compared \System{} with \textit{ProTeGi}~\cite{pryzant2023automatic}, a state-of-the-art automatic prompt optimization library. 
ProTeGi is a general-purpose, non-parametric algorithm that generates performant prompts from user labels.
In detail, it uses random batches of training
examples to produce ``gradients'' in natural language---which are
descriptions of the current prompts’ flaws regarding this batch---and then edit the current prompt in the opposite semantic direction of the gradient.
In each round, it generates and evaluates several candidate improvements, repeating this process for several rounds before selecting the most performant prompt through beam search.

Since ProTeGi was not originally designed for content curation, we adapted it to ensure a fair comparison. First, because the algorithm assumes a task-specific evaluation component, we used our own classification algorithms to estimate performance on the training examples. Second, while ProTeGi assumes an ideal training dataset reflecting task requirements, we instead applied our active learning component to construct such a dataset during initialization.
We implemented this baseline condition following their open-source code on GitHub\footnote{\url{https://github.com/microsoft/LMOps/tree/main/prompt_optimization}} and adopted most of its parameters.
To ensure comparable search budgets across conditions, we aligned beam search settings: each condition could generate at most four prompt iterations and maintain a beam size of two.
Finally, both \System{} and the baseline were implemented in a similar web interface for consistency.

\subsubsection{\textbf{Study Procedures}}
We employed a within-subjects design with the two experiment conditions described above.

\textbf{Stage 1: Study Onboarding.}
We started the experiment by briefing the participants and warning them of the possibility of encountering profanity and hate speech.
After obtaining their informed consent, we asked participants to envision throughout the study that they were managing YouTube comments and presented them with a choice between two content classification tasks: political comments from \textit{TheYoungTurks} channel, or food comments from \textit{BestEverFoodReviewShow} channel. 
Eight participants chose each dataset. 

\textbf{Stage 2: Ground-Truth Labeling.}
Participants then labeled comments from the three datasets listed in Section \ref{datasets} as ``Catch'' or ``Not Catch'' to indicate whether each comment matches their interpretation of the target description. 
Since previous research suggests that users may label examples inconsistently, thus harming algorithmic training ~\cite{kulesza2014structured}, we emphasized that participants should try to label comments in a consistent manner.

\textbf{Stage 3: Initialization of Filters.}
The 20 training comments that participants labeled were used to generate optimized prompts using the two experimental conditions.
We asked participants to review the two resulting filters, with a counterbalanced design helping counter any potential learning effects. 
For each filter, participants first read its prompt, then reviewed its decisions on comments from the training and audit dataset.
They then reported their perceptions of each filter in a survey (more details in Section \ref{sec:evalmetrics}).

\textbf{Stage 4: Iteration on Filters.}
Next, participants were asked to iterate on their filter using the methods in the \System{} condition and to review the automatically refined filter in the baseline condition, with the order of these conditions randomized. 
In the \System{} condition, participants refined their filter three times using whichever combination of fixing one mistake, reviewing common failure patterns, or directly editing the overall description that they preferred, with the goal of best aligning the filter to their classification preferences.
In the baseline condition, we ran the automatic prompt optimization algorithm for three rounds on the same initialized filter.
We did not use the filter initialized in the baseline condition here so that we could focus on measuring the effect of human interventions against automatic iterations. 
Participants then reviewed the optimized prompt and skimmed the decisions that the filter made.
After finishing both conditions, participants reported their perceptions of each approach (see Section \ref{sec:evalmetrics}).

\textbf{Stage 5: Semi-Structured Interviews.}
Finally, we asked participants various questions to understand their perception of the initialized filters and experience with iterating on their filters. Examples of these questions include ``\textit{Which filter's overall description do you prefer, and why?}'' and ``\textit{What do you like or dislike about the fix one mistake feature?}'' We also inquired whether there were any additional features that participants might have liked in the iteration process.

\subsubsection{\textbf{Evaluation Metrics}} \label{sec:evalmetrics}
We use the following means to evaluate filter initialization and iteration.

We measured the performance of the resultant filters from the initialization stage and the iteration stage on the test dataset that each participant labeled.
We further gathered participants’ subjective perceptions using a five-point Likert item (from Strongly Disagree to Strongly Agree) about the following measures:
\begin{itemize}
    \item \textit{Performance Perception}. To measure participants' perception of filter performance, we asked participants to rate the level of performance with which each condition's filter classified comments in the initialization stage, as well as how effectively the iteration approach in each condition improved their filter.
    \item \textit{Interpretability}. To assess the extent to which participants understood the decisions of the filters at the initialization stage, we asked participants to rate how easily they understood both the overall description of each filter and why each filter caught or did not catch comments.
    Regarding the iteration stage, we asked participants to rate how easily they understood both why \System{} suggested its iterations and why the baseline condition's approach made its iterations.
    \item \textit{Task Load}. For the iteration stage, we also adopted two applicable items from the NASA-TLX survey~\cite{hart1988development} regarding mental demand and effort, excluding irrelevant questions.
    We also asked participants' preferences about iteration approaches in the end of experiments.
\end{itemize}

\begin{figure*}[t]
    \centering
    \begin{subfigure}[t]{\textwidth}
        \centering
        \includegraphics[width=0.85\linewidth]{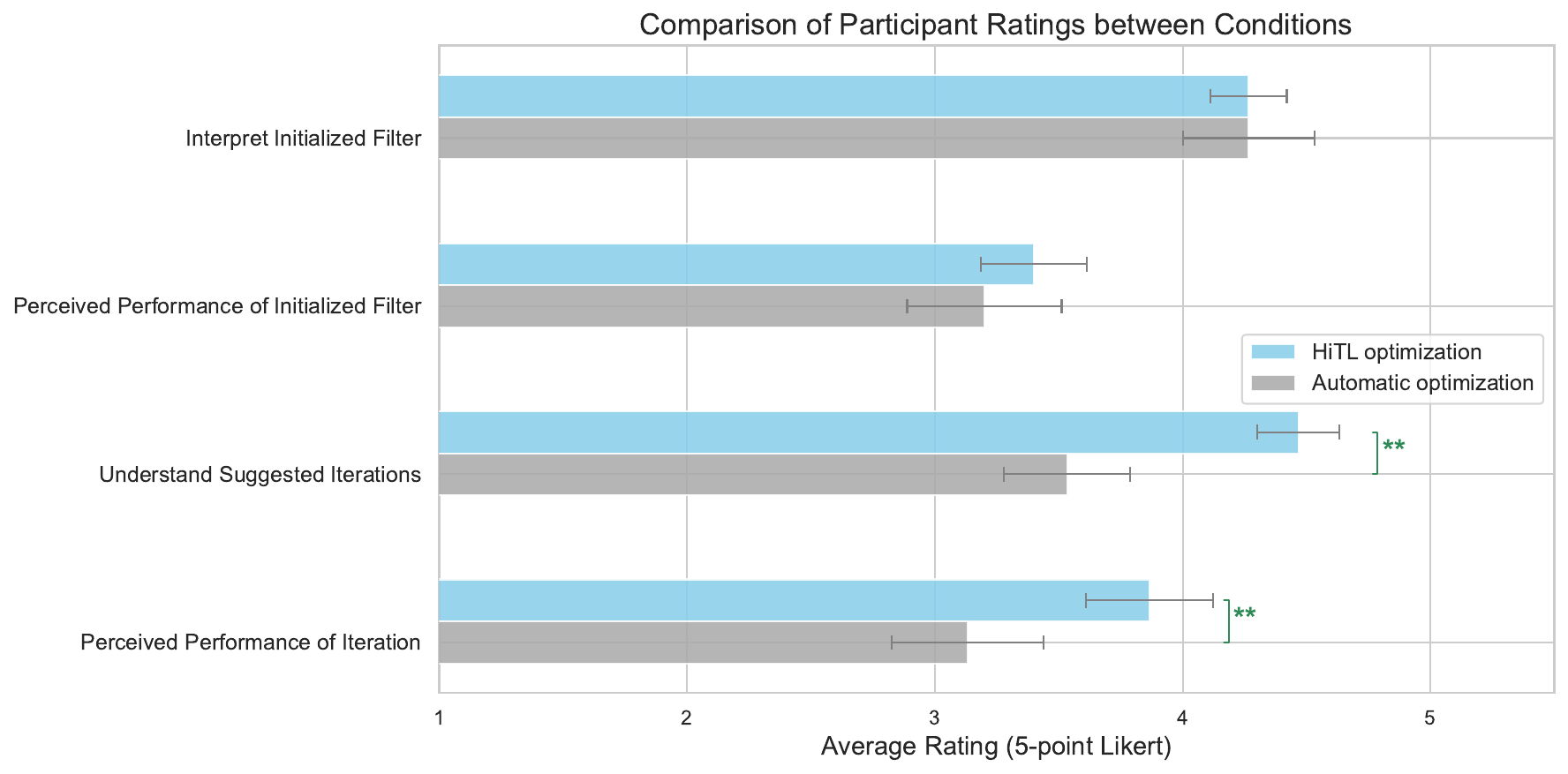}
        \label{fig:experiment-comparison-shared}
        \caption{
        At the end of the initialization and iteration stages, participants evaluated the interpretability and perceived performance of the filters generated by the two optimization approaches.
        Error bars show standard errors; significance markers indicate paired comparisons. 
        We found no significant differences at initialization. 
        By contrast, during the iteration stage, participants rated \System{} filters as both more interpretable and higher performing. 
        This suggests that automatic prompt optimization produces less interpretable edits as iterations accumulate.}
    \end{subfigure}
    \vspace{1em} 

    \begin{subfigure}[t]{\textwidth}
        \centering
        \includegraphics[width=0.85\linewidth]{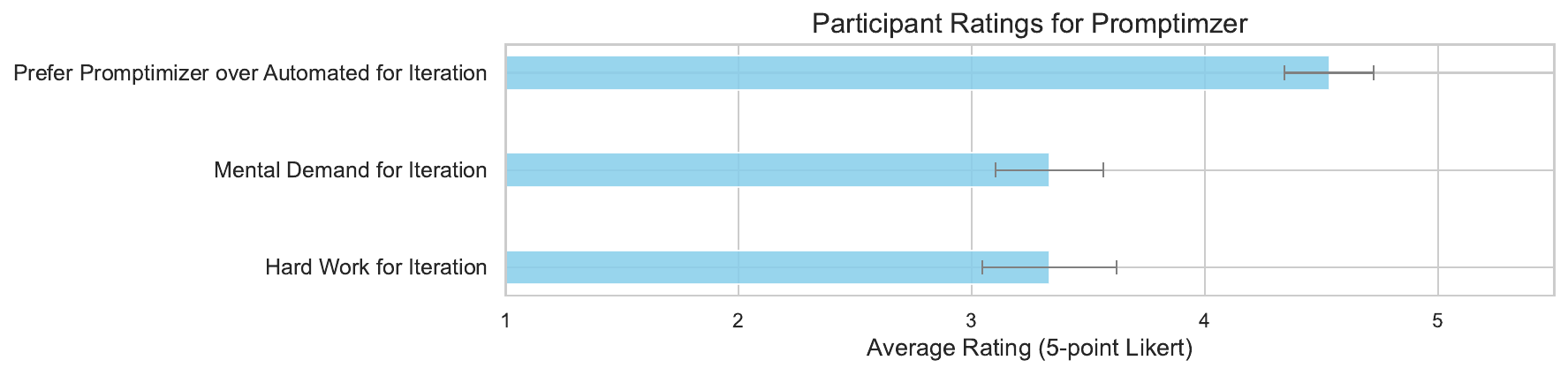}
        \label{fig:experiment-comparison-only}
        \caption{We also measured the task load of manual iteration for \System{} and asked whether participants preferred iterating with it compared to automatic optimization. We found that participants unanimously preferred \System{} for iterations, although they reported that iterating with \System{} was fairly mentally demanding.}
    \end{subfigure}
    \caption{\textbf{Participant Ratings for \System{} and Automatic Optimization.}}
    \label{fig:experiment-comparison}
\end{figure*}

\begin{table*}[t]
\small
\centering
\captionsetup[table]{skip=10pt}
\begin{tabular}{l|l|cc|cc|cc|cc}
\toprule
\multirow{2}{*}{\textbf{Comparison Groups}} & \multirow{2}{*}{\textbf{Condition}} 
& \multicolumn{2}{c|}{\textbf{Accuracy}} 
& \multicolumn{2}{c|}{\textbf{Precision}} 
& \multicolumn{2}{c|}{\textbf{Recall}} 
& \multicolumn{2}{c}{\textbf{$F_1$ Score}} \\
& & Est. & Std. Err. & Est. & Std. Err. & Est. & Std. Err. & Est. & Std. Err. \\
\midrule
\multirow{2}{*}{\textbf{After Initialization}} 
& \System{} &  0.570 & 0.114 & 0.711* & 0.213 &  0.507 & 0.117 &  0.710 & 0.111\\
& Baseline  & 0.604 & 0.140 & 0.652  & 0.231 & 0.619*** & 0.147 & 0.707 & 0.090\\
\midrule\midrule
\multirow{2}{*}{\textbf{After all Iterations}} 
& \System{} &  0.638 & 0.128 & 0.717 & 0.210 &  0.599   & 0.091 &  0.756 & 0.066\\
& Baseline  & 0.630 & 0.130 & 0.732  & 0.215 & 0.598 & 0.154 & 0.749 & 0.084 \\
\midrule\midrule
\textbf{\System{}}
& After Initialization &  0.570 & 0.114 & 0.711 & 0.213 &  0.507 & 0.117 &  0.710 & 0.111\\
\textbf{Within} & After all Iterations     & 0.638** & 0.128 & 0.717 & 0.210 &  0.599**   & 0.091 &  0.756** & 0.066\\
\bottomrule
\end{tabular}
\caption{\textbf{Post-hoc paired t-test results for classifier performance on the test dataset ($n=16$).} 
We report pairwise differences in performance metrics between \System{} and automatic prompt optimization at both initialization and iteration, and show how three rounds of \System{} iterations contributed to performance gains. 
Significance: $p<.001$~(***), $p<.01$~(**), $p<.05$~(*).}
\label{tab:pairwise-test-results}
\end{table*}

\subsection{Data Analysis}

\subsubsection{Quantitative Analysis}
For our quantitative analysis, we used paired samples $t$-tests to compare participants’ subjective ratings of filter interpretability and perceived performance across the two conditions. Beyond subjective perception of performance, we also evaluate objective classifier performance on the held-out test dataset, including accuracy, precision, recall, and $F_{1}$ score, and also compare them with a paired $t$-test.

\subsubsection{Qualitative Coding}\label{qualitative-analysis}
Our qualitative data comprised semi-structured interview data and responses to open-ended questions in surveys. We employed a reflexive thematic analysis approach~\cite{Braun2019} to explore participants' experiences and challenges in initializing and improving personalized classifiers with \System{}.
Reflexive thematic analysis has been widely used in HCI research to understand users’ experiences and views, as well as factors that influence particular phenomena or processes~\cite{Braun2019}. During data collection, the first author took detailed debriefing notes after each interview to document emerging themes. The authors then collectively reviewed the debrief notes and discussed themes in weekly group meetings. Recordings were automatically transcribed into text. The first author then open-coded the data on a line-by-line basis, and the remaining authors reviewed the transcripts and added codes. 368 codes were generated from the open-coding process. The authors clustered the open codes into high-level themes in a codebook and iteratively improved the codebook through discussion. Example codes are ``\textit{Trade-offs between performance and agency}'' and ``\textit{Iterations: preferred ways of iterations}.'' Finally, the authors applied the codes to the data to complete the thematic analysis.

\subsection{Results}

\begin{figure*}[t!]
    \centering
    \includegraphics[width=.99\textwidth]{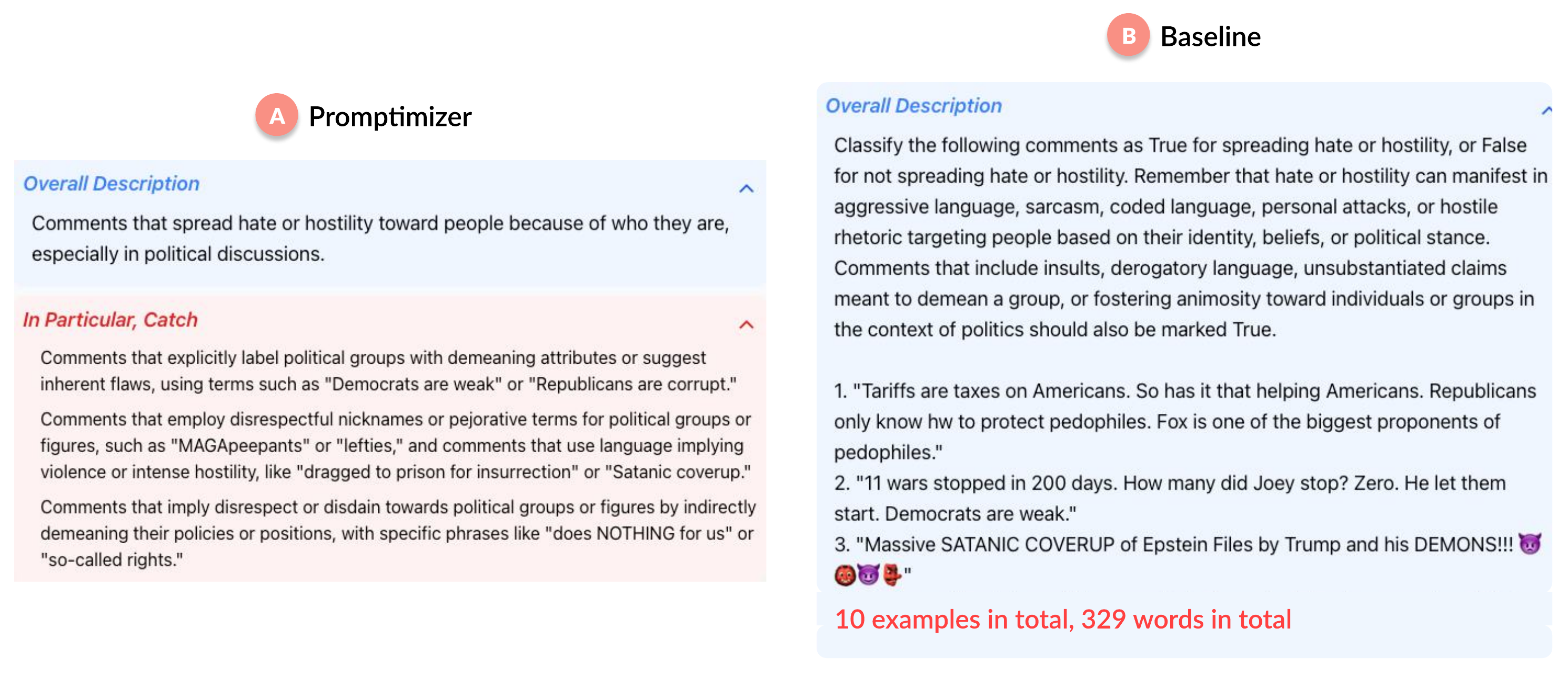}
    \caption{\textbf{Initial Prompts from the Two Experimental Conditions}. \textbf{(A)} This is the prompt after three iterations in \System{} for one participant, with the overall description, and three positive rubrics. \textbf{(B)} The corresponding prompt after three rounds of iterations by the baseline APO condition is a freestyle string of text, which simply chose to add more examples as iterations.}
    \hfill
    \label{fig:prompts}
\end{figure*}

\subsubsection{\textbf{Users perceived \System{} as generating significantly more interpretable iterations than the automatic condition (see Figure \ref{fig:experiment-comparison}(a))}}
The structured prompts in \System{} helped participants more easily understand filter decisions.
For reference, Figure \ref{fig:prompts} displays initial prompts generated in the two experimental conditions for one participant.
As E13 explained, ``\textit{I think having it structured with the overall description and then rubrics for `catch' and `do not catch' is really helpful because they emphasize some of the gray-area or edge cases. I think it helps me be more certain in my decisions, and also helps the filter itself make those decisions perfect.}''
Similarly, E15 found automated prompts less interpretable due to their lack of structure: ``\textit{I didn't really like this huge block of text [in the automated condition], and I found it a little bit hard to follow.}''
Participants also emphasized that the structured format of \System{} supported more targeted iterations.
For instance, E16 noted, ``\textit{It felt useful that, in the interactive condition, the description of what to catch was kept separate from the description of what not to catch, because it made them into discrete standards that I could iterate on. }''
In addition, because participants directly influenced the direction of prompt changes, they reported a stronger sense of the reason behind each iteration. As E3 reflected, ``\textit{When you don't have the involvement, it will give these vague explanations that you might not necessarily agree with, whereas in the interactive approach, you can have more inputs back and forth.}''
Despite these benefits in the iteration stage, we did not find evidence that \System{} produced more interpretable prompts during the initialization stage (see Figure~\ref{fig:experiment-comparison}). This is likely because prompts at both conditions had not yet grown complex. 
As a result, participants could easily interpret classification outcomes in both conditions at this early stage.


\subsubsection{\textbf{The performance of initialized and iterated filters within each condition were comparable (see Table \ref{tab:pairwise-test-results}).}}
To support interpretability, \System{} constrained the search space of prompt optimization and incorporated multiple forms of user guidance. Importantly, these benefits did not come at the cost of performance: on participants’ test datasets, both conditions achieved comparable performance.
However, participants often perceived \System{} as producing more performant iterations. 
On one hand, it might be because that being involved in the iteration process helped them identify and correct mistakes that they cared about most, even if such improvements were not fully reflected in a small test dataset of only 100 comments.
Nevertheless, we cannot rule out the possibility that these perceptions were shaped by a few salient examples rather than systematic differences, or users were biased based on their efforts.

\subsubsection{\textbf{Users unanimously preferred \System{} for iterating on their filters.}}
As Figure \ref{fig:experiment-comparison} indicates, all participants preferred \System{} over automatic optimization, mostly because they enjoyed having control in the classification process despite acknowledging its mental demand.
Many criticized the automatic condition for relying solely on their initial labels, which provided limited guidance.
In contrast, participants could exert more control in the \System{} condition, where they could pick mistakes that they wanted to fix, manually clarify their preferences, and select which iterations they preferred.
Consequently, participants found that the automated approach failed to incorporate nuances of their preferences. For instance, E7 explained, ``\textit{I preferred interactive because it allows the niche differences to be applied to the filters.}''
Participants also noted that automatic optimization sometimes reinforced outdated labels, whereas iteration in \System{} let them revise their views.
For instance, E16 updated their labels for some comments when iterating on \System{}: ``\textit{I was able to easily switch from `catch' to `not catch'...when I realized I really should not have caught this comment.}''
\System{} also enabled participants to better navigate trade-offs between precision and recall, as they were able to choose between addressing false positives and false negatives.
E16 found this an important part of his preferences: ``\textit{I'm inclined to prefer under-filtering [high recall] over over-filtering [high precision]...I prefer the idea of having the option to exercise that control over what I myself am seeing in specific ways.}''


\textbf{Additionally, \System{} supported participants in pursuing their varied preferences about how they wanted to make iterations.}
Many participants reported that fixing one mistake allowed them to prioritize certain mistakes and thus steer their filters in a particular direction.
For instance, for the cultural food experience filter, E1 ``\textit{wanted more specific food comments, so [he] just picked those mistakes and hit refine.}''
Similarly, E8 wanted to prioritize mistakes about xenophobia and racism from the political hate speech filter over comments that were simply mean.
Iteration based on one mistake also allowed participants to incorporate more nuances into their filters, since they could clarify their preferences.
As E3 said, ``\textit{It's nice that I could type and tell the LLM, `Oh, well, I didn't exclude these comments because of this, or I did exclude these comments because of this.'}''
According to participants, the automatically generated list of clarification candidates often aligned with their preferences. 
These participants also feared that common failure patterns might result in vague, overly broad filtering.
E7 prioritized precision over recall and thus elaborated on this: ``\textit{[Fixing one mistake] is more easy for me to visualize what comments it will potentially exclude, [whereas for common failure patterns], I don't have as much control...I don't know exactly what criteria are used.}''



Meanwhile, other participants preferred iterating on their filters using common failure patterns.
They found that these failure patterns helped generalize beyond details in a single mistake and surfaced important iterations.
For instance, E14 reported, ``\textit{I personally like the common failure patterns because they weren't as general as the automated approach but it was general enough to stimulate my thinking...Fixing one mistake in the grand scheme of things will not generalize to multiple mistakes.}''
Participants felt that this feature would be especially useful if there were many mistakes.
E3 explained that ``\textit{when there are 10 mistakes, I have the ability to go in and fine-tune, but if it was one thousand comments, then I would just want a better failure pattern to recognize the commonalities.}''
In fact, these participants often found it overwhelming to pick a particular mistake to fix with the fix-one-mistake feature, as E9 mentioned: ``\textit{There was not one particular comment that stood out to me as a good use case of changing, so I will just pick the first mistake I see.}'' They preferred common failure patterns automatically picking mistakes with commonalities through clustering.
Additionally, participants proposed new ways of exploring failure patterns that lie in between these two options, such as manually selecting a set of mistakes and then collectively clarifying those preferences. 




\section{\textsc{\textsc{\YouTubeSystem{}}}: A System for Content Creators to Create Comment Filters}

\begin{figure*}[t!]
    \centering
    \includegraphics[width=.9\textwidth]{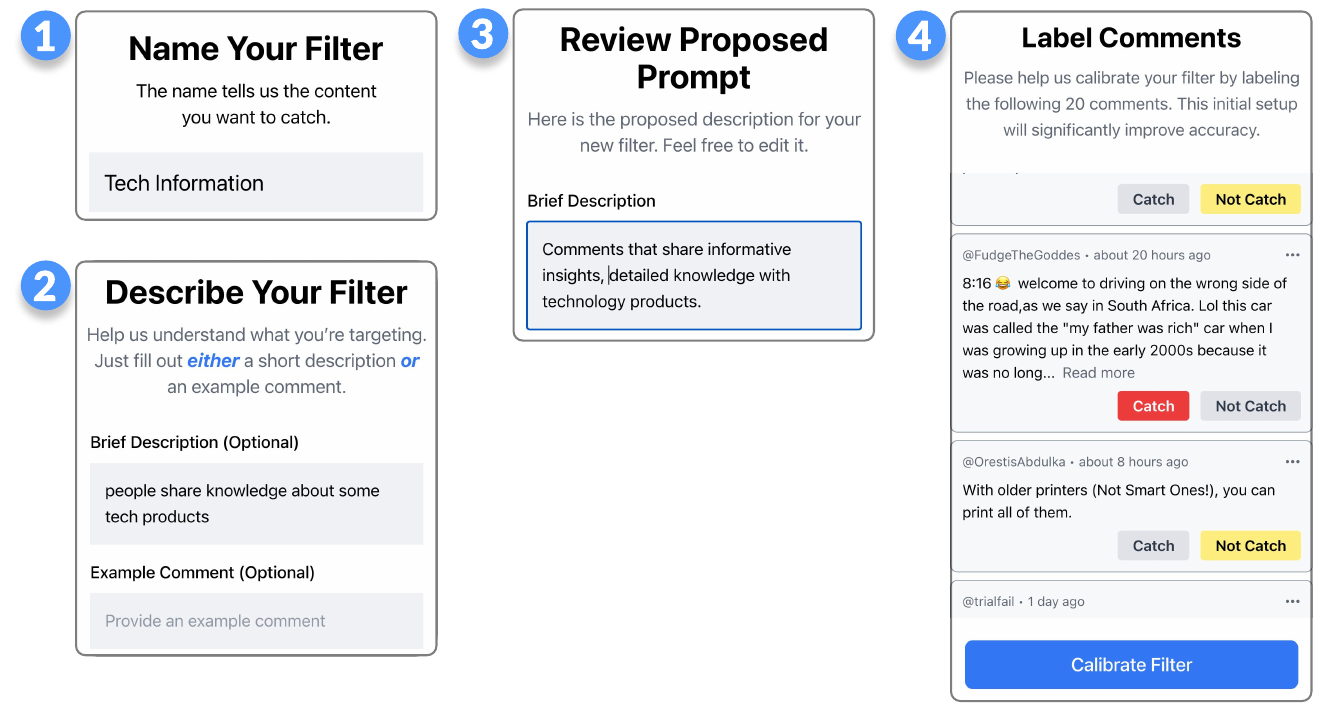}
    \caption{\textbf{\textsc{\YouTubeSystem{}} Initialization}. To initialize filters, users begin by naming them \textbf{(1)}, then provide a short description or example comment \textbf{(2)}. These inputs allow \textsc{\textsc{\YouTubeSystem{}}} to generate a more detailed draft description of users' preferences, which is open to users' review and edits \textbf{(3)}. Since users find it easy to communicate nuances in their preferences by labeling examples~\cite{wang2025end}, we surface 20 ``interesting'' examples and ask for users' labels \textbf{(4)}. 
    We then automatically optimize the initial description to better reflect users' preferences.}
    \hfill
    \label{initialize}
\end{figure*}

We implemented our human-in-the-loop prompt optimization workflow into \textsc{\YouTubeSystem{}}, a system that supports YouTube content creators in creating custom filters for curating and moderating comments on their videos.
We chose this as our test setting because content creators have a clear need for easily authoring personal content filters.
They face risks of hate and harassment, making protection from unwanted comments especially important~\cite{thomas2022s, jhaver2022designing}.
Simultaneously, they rely on curating relevant comments to better engage with and understand their audiences~\cite{choi2025proxona}.
While high-profile creators may have staff support for comment management, most creators dedicate their time to video production and have limited expertise or bandwidth for content classification~\cite{weber2021s}.
In implementing \textsc{\YouTubeSystem{}}, we conducted pilot studies through two group sessions with our lab members and four YouTube content creators. These sessions helped us iteratively develop a user-friendly system tailored for creators.

\begin{figure*}
    \centering
    \includegraphics[width=\textwidth]{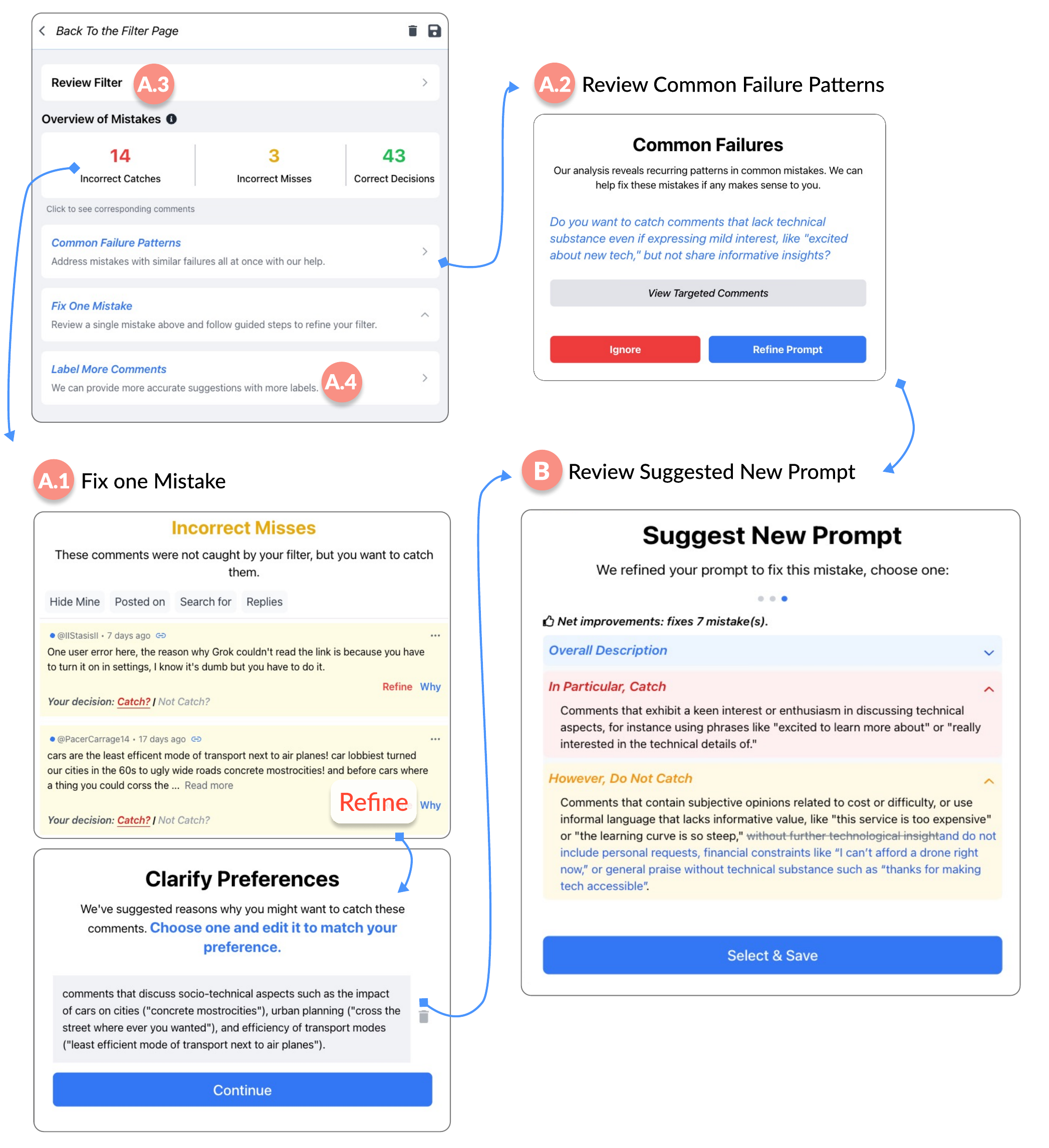}
    \caption{\textbf{\textsc{\YouTubeSystem{}} Iteration}. \textsc{\textsc{\YouTubeSystem{}}} supports users in making meaningful iterations on their filters in four ways. \textbf{(A.1) Fix One Mistake}. \textsc{\textsc{\YouTubeSystem{}}} asks users to clarify their rationale by providing a few candidates that users can select and edit. Based on users’ clarifications, \textsc{\textsc{\YouTubeSystem{}}} generates a set of revised prompts for review, allowing users to select the most suitable option, seen in (B). 
    \textbf{(A.2) Review Common Failure Patterns}. \textsc{\textsc{\YouTubeSystem{}}} also summarizes common failure patterns for users' review, followed by new prompt suggestions (B).
    \textbf{(A.3) Manual Edits}. Additionally, users can modify their filter descriptions directly. \textbf{(A.4) Label More Comments.} Users can label additional comments, similar to the initialization stage.}
    \hfill
    \label{iterate}
\end{figure*}

\subsection{System Walkthrough}

We illustrate the workflow of \textsc{\YouTubeSystem{}} through a walkthrough from the perspective of \Username{}, a content creator who wants to better understand her audience’s opinions on different technological products.

\subsubsection{Initialization}
\begin{figure*}[t!]
    \centering
    \includegraphics[width=\textwidth]{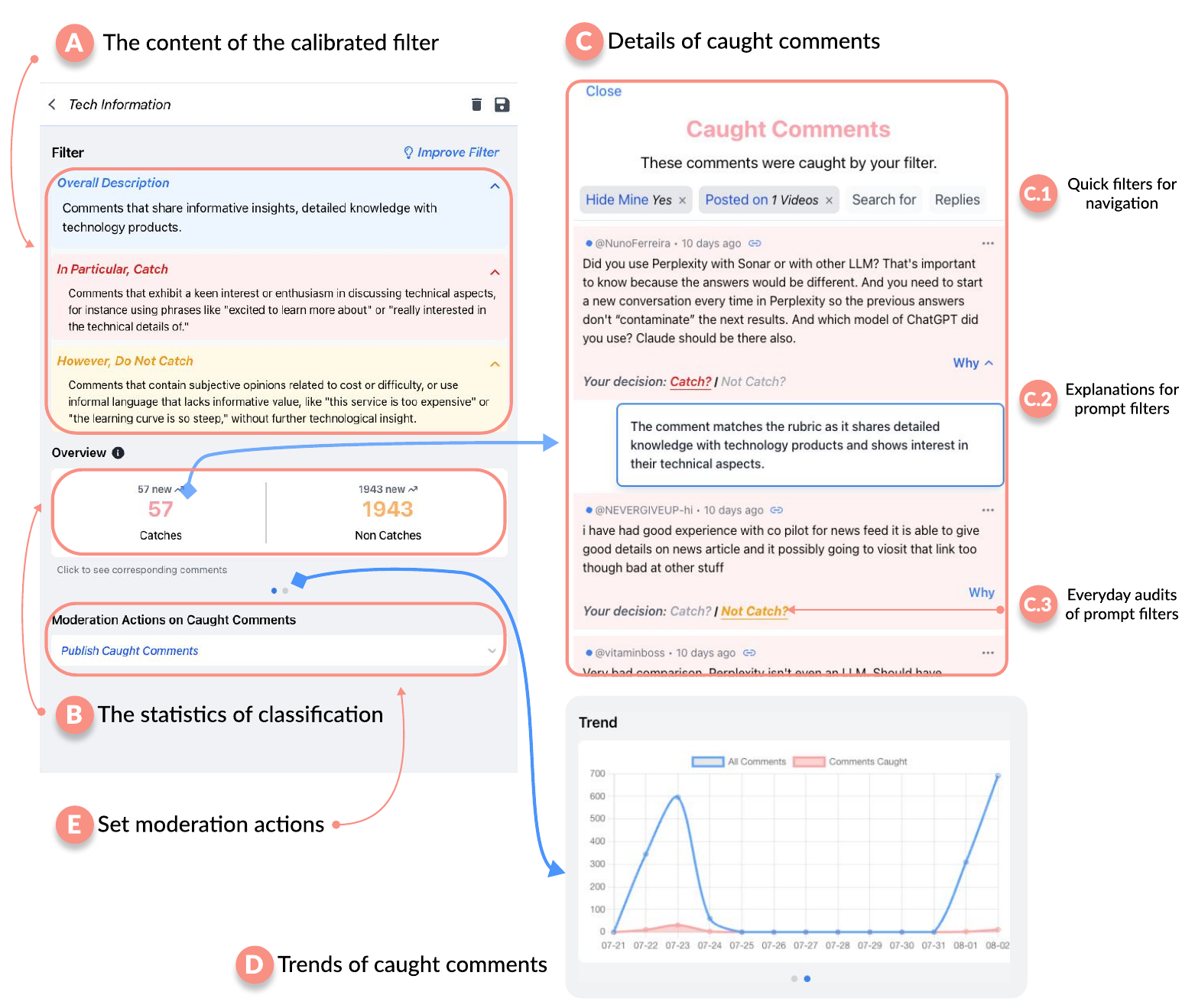}
    \caption{\textbf{\textsc{\YouTubeSystem{}} Audits}. After users calibrate their initial descriptions with labels, they can review their filter, deciding whether they want to apply it to their comment sections. \textbf{(A)} The filter content is displayed, including an overall description plus positive and negative rubrics. \textbf{(B)} By applying this filter to their comments, users can review statistics on how many comments were caught or not. \textbf{(C)} Clicking on these numbers opens a sheet displaying the corresponding comments. We offer quick filters for users to more easily navigate these comments (C.1), provide explanations for predictions (C.2), and allow users to audit their filters' decisions, enabling future improvements (C.3). \textbf{(D)} In the carousel on the main page, users can switch to a chart showing trends in the number of caught comments over time. \textbf{(E)} Users can assign moderation actions to caught comments, such as publishing or holding them for review, if they grant \textsc{\YouTubeSystem{}} the relevant YouTube permissions.}
    \hfill
    \label{audit}
\end{figure*}

Figure~\ref{initialize} shows how \Username{} can initialize a personalized content filter. She begins by naming the filter and describing her preferences. If she finds this difficult, she can instead provide example comments that reflect which content she wants to catch.
These inputs allow \textsc{\YouTubeSystem{}} to generate a more detailed description of her preferences, which remains open to further edits: ``\textit{Comments that share informative insights, detailed knowledge with technology products.}''
However, this description may not fully capture \Username{}'s nuanced preferences~\cite{wang2025end, kumar2023watch}. For example, \Username{} considers electric vehicles as technology products, while the LLM might not; similarly, terms like ``insights'' or ``detailed'' may be interpreted differently. 
To communicate these nuances, \textsc{\YouTubeSystem{}} asks \Username{} to review and label 20 ``interesting'' comments selected using active learning. These labels allow \textsc{\YouTubeSystem{}} to automatically optimize the initial description so that it better reflects \Username{}'s preferences.
\Username{} can then review her initialized filter (Figure~\ref{audit}), with positive and negative rubrics added according to her initial labels.
For example, \textsc{\YouTubeSystem{}} infers that \Username{} does not want to catch comments that only mention cost or difficulty. 
Satisfied with the description, \Username{} deploys the filter.

\subsubsection{Iteration}
As \Username{} continues to use \textsc{\YouTubeSystem{}}, she encounters new mistakes and seeks to make improvements. 
As shown in Figure \ref{iterate}, the main page of \textsc{\YouTubeSystem{}} summarizes the number of false positives, false negatives, and correct predictions from all comments she has annotated so far. 
It offers three primary approaches for refinement.

\textbf{Fix One Mistake}.
While reviewing mistakes, \Username{} may find one that feels especially important. 
\textsc{\YouTubeSystem{}} prompts her to clarify her intent in more detail. 
To reduce effort, \textsc{\YouTubeSystem{}} provides several LLM-generated candidate rationales that \Username{} can select or edit (Figure~\ref{iterate} A.1).
Based on this clarification, \textsc{\YouTubeSystem{}} creates new prompt suggestions targeted at fixing the mistake (Figure~\ref{iterate} B).
Each suggestion highlights the proposed changes and estimates how many mistakes it resolves or introduces. 
\Username{} can then review and select her preferred iteration.

\textbf{Review Common Failure Patterns}. 
For \Username{}, reviewing mistakes individually can be overwhelming, especially when comments are distressing or no single mistake feels especially important. 
To help, \textsc{\YouTubeSystem{}} clusters mistakes into common failure patterns and highlights those that occur most frequently (Figure~\ref{iterate} A.2). 
\Username{} can examine these patterns and, if she selects one, she can review new prompt suggestions generated by \textsc{\YouTubeSystem{}} (Figure~\ref{iterate} B), similar to the ``Fix One Mistake'' workflow. Unlike that workflow, however, explicit rationales are not requested, since \Username{}'s preferences are inferred from the clustered mistakes.

\textbf{Make Manual Edits}. \textsc{\YouTubeSystem{}} also allows \Username{} to edit her filter description directly (Figure~\ref{iterate} A.3). However, since manual prompt engineering can be difficult, we keep this option secondary.

\textbf{Label More Comments}. Finally, \Username{} might wish to refine her filter but has seen few mistakes so far. 
\textsc{\YouTubeSystem{}} can surface additional ``interesting'' comments for review (Figure~\ref{iterate} A.4). 
This process uncovers more mistakes while expanding the ground-truth dataset, enabling better evaluation of iterations.

\subsubsection{Audits}
\Username{} can learn more about her filters on the main filter page (Figure~\ref{audit}).
New comments are also automatically fetched from her channel and classified every hour. 
Beyond the top section of filter descriptions, the middle section shows summary statistics of caught and uncaught comments. 
By clicking ``57 catches,'' \Username{} can explore examples, filter them by various criteria, and see explanations for why each comment was included.
She can also audit her filter by labeling whether comments should or should not have been caught. 
Because she already reviews comments as part of her routine, this becomes an ongoing, lightweight audit. 
A toggle also lets her view a line chart of caught comment trends over time.
At the bottom, \Username{} chooses moderation actions for caught comments: \textit{do nothing}, \textit{hold for review}, \textit{delete}, \textit{publish} (if comments are withheld by default), or \textit{reply with a template}. The \textit{do nothing} option supports her when she wants to curate relevant comments only for personal reference.

\subsection{System Implementation}
\textsc{\YouTubeSystem{}}, hosted on a web server, consists of a frontend web interface in React and a Django backend in Python. 
The backend authorizes YouTube accounts, fetches comments, and performs moderation actions via Google YouTube API. 
Throughout our implementation, we used OpenAI's \texttt{gpt-4-1106-preview} model.

\section{Field Study with YouTube Content Creators}
To investigate how people would use \textsc{\YouTubeSystem{}} to classify content in their daily workflow, we conducted a field study with 10 YouTube content creators across three weeks to answer the following research questions:

\begin{itemize}
    \item \textbf{RQ4}: How does \textsc{\YouTubeSystem{}} help content creators manage their comment sections?

    \item \textbf{RQ5}: Does \textsc{\YouTubeSystem{}} support content creators to author interpretable and performant filters in the wild?

    \item \textbf{RQ6}: How do content creators find the \System{} workflow within \textsc{\YouTubeSystem{}} to initialize and iterate on their filters?

\end{itemize}

\subsection{Recruitment and Participants}
\label{field study recruitment}
We recruited 10 active YouTube content creators, labeled P1-P10, primarily through purposive sampling, supplemented by snowball sampling. 
We only selected participants whose channels displayed at least 100 comments to ensure that they had sufficient comments for \textsc{\YouTubeSystem{}} to filter. 
Our recruitment followed three strategies: posting calls for participation on Discord and Reddit groups for YouTube creators, advertising calls for participation through Slack, and directly contacting YouTubers via LinkedIn messages and cold emails.
Throughout recruitment, we used a pre-screening survey to collect interest and help verify eligibility and credibility, asking about individuals' channel ownership, content topics, and any existing comment management practices.
We discussed difficulties in our recruitment process and call for more future research to investigate better practices in the Appendix \ref{discussion:recruitment}.

Table~\ref{tab:fieldstudy} summarizes participant demographics. To preserve privacy, we conceal names and channel details. All participants engaged with \textsc{\YouTubeSystem{}} over three weeks, with an average total involvement of three hours each. They received a \$100 gift card as compensation. This study was reviewed by our university IRB and deemed exempt.

\begin{table}[h!]
  \caption{\textbf{Participant Information.} Demographic information of field-study participants, who were active YouTube content creators at the time of the study. ``Topics'' refers to the main topical focuses of participants' channels. ``Recruitment'' indicates how participants were recruited.}
  \label{tab:fieldstudy}
  \begin{tabular}{cccccc}
    \toprule
    \textbf{Participant} & \textbf{Gender} & \textbf{Subscriber Count} & \textbf{Video Count} & \textbf{Topics} & \textbf{Recruitment} \\
    \midrule
    P1 & Male & 100k-200k & 500-1k & Gaming & Slack \\
    P2 & Male & 10k-20k & 60-70k & Science & Discord \\
    P3 & Male & 10k-20k & < 500 & Culture, Science, Self-Help & Slack \\
    P4 & Female & > 1m & < 500 & Lifestyle & Email \\
    P5 & Male & < 500 & < 500 & Gaming & Discord \\
    P6 & Female & 10k-20k & 1k-10k & Culture & Snowball Sampling \\
    P7 & Female & 500-1k & < 500 & Dance, Lifestyle & Email \\
    P8 & Male & < 500 & < 500 & Teaching & Discord \\
    P9 & Male & 10k-20k & < 500 & LGBTQ+, Lifestyle & Email \\
    P10 & Male & 300k-400k & < 500 & Disability, Lifestyle & Email \\
    \bottomrule
  \end{tabular}
\end{table}

\subsection{\textbf{Study Procedure}}\hfill

\textbf{Stage 1: Study Onboarding.}
We began by briefing participants and walking them through \textsc{\YouTubeSystem{}}’s functionalities over Zoom. Participants were asked to enter their YouTube username, which loaded their comments into \textsc{\YouTubeSystem{}}. They then tried each feature and raised any questions. We concluded by explaining the field-study procedure and their weekly tasks, emphasizing that participants could stop the study whenever they wanted. We gained their explicit consent before proceeding. 

\textbf{Stage 2: System Use across Three Weeks.}
Over a three-week period, participants completed the same weekly tasks: creating at least one new filter, reviewing caught comments every other day to mark some mistakes, and iterating on their filters at least three times using the provided strategies. 
We confirmed this activity by verifying it in our user logs. 
Since each participant entered the username for their personal YouTube account, we let them decide individually whether they wanted \textsc{\textsc{\YouTubeSystem{}}}'s filtering actions to apply to their account's comments. 
Participants also completed a weekly survey about their experience using \textsc{\textsc{\YouTubeSystem{}}}. 
It included two items from the Usability Metric for User Experience (UMUX) survey~\cite{lewis2018measuring} to measure system usability for creating and iterating filters in \textsc{\textsc{\YouTubeSystem{}}}, as well as \textsc{\textsc{\YouTubeSystem{}}}'s general utility for content curation.
Participants also rated the interpretability of \textsc{\YouTubeSystem{}} and provided suggestions for improvement in an open-ended question.

\textbf{Stage 3: Semi-Structured Interviews }
After the three weeks, we conducted semi-structured interviews with each participant over Zoom. They averaged 87 minutes. Interviews explored participants' experiences with creating and iterating on filters. Example questions include, ``\textit{What kinds of filters did you create?}'' and ``\textit{How do you feel about the experience of creating a filter in our system?}''
After their interviews, participants also completed an end-of-study survey with all 10 SUS items~\cite{brooke1996sus} to measure the overall usability of \textsc{\textsc{\YouTubeSystem{}}}.

\subsection{Data Analysis}
For qualitative data, we applied the same analysis procedure used in our first study (see Section~\ref{qualitative-analysis}). 
Example codes are ``\textit{Curation for engaging with audience}'' and ``\textit{Iteration: clarify preferences.}''
Because the field study did not include a baseline condition for comparison, we did not conduct statistical tests on the survey data.

\subsection{Results}


\begin{table}[t]
\small
\centering
\setlength{\tabcolsep}{6pt}
\renewcommand{\arraystretch}{1.1}
\begin{tabular}{p{.48\linewidth} p{.48\linewidth}}
\toprule
\multicolumn{2}{@{}l}{\textbf{Content Moderation (23 filters)}}\\
\midrule
\begin{tabular}[t]{@{}l@{}}
Spam (×3)\\
Hate speech (×3) \\
Personal attacks (x2) \\
Defamation \& slander (×2)\\
Inappropriate tone (×2)\\
Scams (×2)\\
Violence threats\\
Political agenda pushing\\
\end{tabular}
&
\begin{tabular}[t]{@{}l@{}}
Rude or insensitive comments\\
Sarcasm\\
Promotional comments\\
Self-promotion\\
Comments negatively comparing past relationships\\
Romantic relationship speculation\\
Irrelevant comments
\end{tabular}
\\[4pt]
\toprule
\multicolumn{2}{@{}l}{\textbf{Content Curation (18 filters)}}\\
\midrule
\begin{tabular}[t]{@{}l@{}}
Constructive feedback or suggestions (×4)\\
Potential video ideas (×2)\\
Comments that deserve an answer or thank you (x2)\\
Questions about spinal cord injury or wheelchair life\\
Questions about featured products\\
Supportive comments\\
Sharing memories\\
\end{tabular}
&
\begin{tabular}[t]{@{}l@{}}
Sharing milestones\\
Sharing information\\
Mentioning Puerto Rico\\
Music-related comments\\
Adaptive tools and resources\\
Mind-body techniques\\
\end{tabular}
\\
\bottomrule
\end{tabular}
\caption{\textbf{Content Filters Authored by Field Study Participants.} We group filters by intent into \emph{moderation} and \emph{curation}. Similar filters are merged with counts in parentheses. Private creator details were removed from filter names.}
\label{tab:filters-moderation-curation}
\end{table}

\subsubsection{\textbf{How does \textsc{\YouTubeSystem{}} help content creators manage their comment sections?}} 
\begin{figure*}
    \centering
    \includegraphics[width=0.7\textwidth]{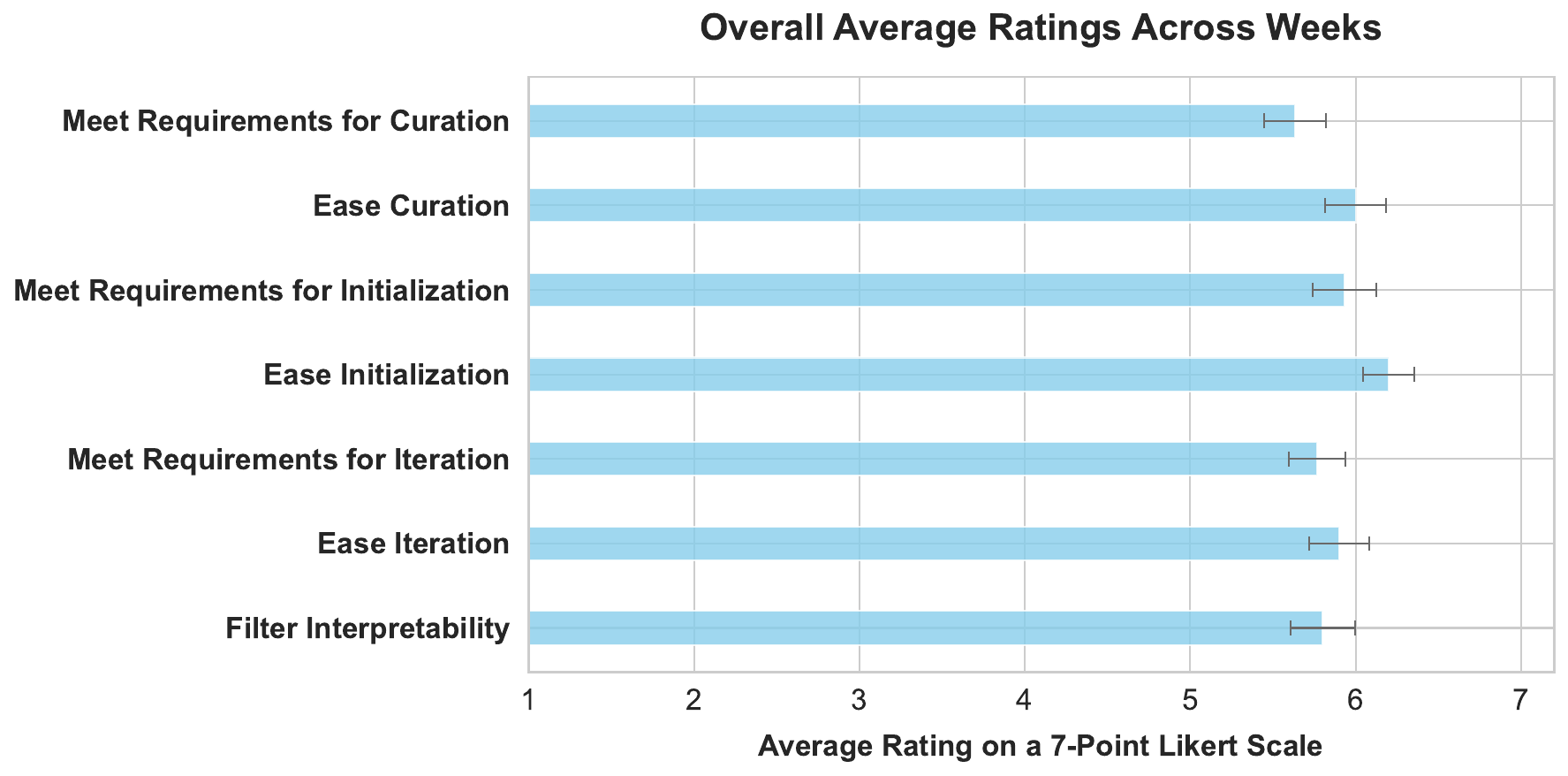}
    \caption{\textbf{Participant Ratings for \textsc{\YouTubeSystem{}}.} 
    We report participants’ average ratings across three weeks on \textsc{\YouTubeSystem{}}'s ease of use, interpretability, and support for content curation, filter initialization, and filter iteration. 
    We do not report weekly results because we did not observe significant differences across weeks.
    Error bars show standard error, with the x-axis truncated to 4–7 for readability. 
    Overall, participants agreed that the system helped them curate content, generate interpretable filters, and easily initialize and refine filters.}
    \hfill
    \label{usability}
\end{figure*}

Participants found that \textsc{\textsc{\YouTubeSystem{}}} helped them easily remove unwanted comments and curate interesting comments for review (see Figure \ref{usability}).
As shown in Table \ref{tab:filters-moderation-curation}, participants created a total of 41 filters with \textsc{\textsc{\YouTubeSystem{}}}, averaging 4.1 filters per person. Their filters can be broadly categorized into content moderation and content curation.

\textbf{Content Moderation.} 
Most participants started by creating filters to identify spam, scams, or hate speech, removing the need to manually read through and discover unwanted comments.
While YouTube allows participants to create a list of blocked words or adjust their sensitivity to inappropriate comments, participants found that \textsc{\YouTubeSystem{}} supported more niche moderation preferences.
For instance, P2 runs a channel that covers natural disasters and created a filter that ``\textit{keeps track of people that push agendas or conspiracy theories.}''
He said, ``\textit{It's definitely helpful for niche specific comments, because a gaming or a vlogging channel likely will not get the kind of comments I get,}'' adding that ``\textit{this is actually the first one of its kind, because with YouTube's built-in stuff, you just pretty much filter basic common negative comments.}''
As another example, P4 created a filter that detects sarcastic comments and another for comments promoting commercial products.

\textbf{Content Curation.} As participants continued using \textsc{\YouTubeSystem{}}, many began to curate interesting comments to better engage with and understand their audience.
A few created filters to learn new content ideas, or to collect constructive feedback from their audiences so that they could create better content.
P7 mentioned, ``\textit{I was sort of seeing what the comments would say for constructive criticism and the suggestions, so I can apply that criticism to any future videos.}''
P6 used \textsc{\YouTubeSystem{}} to identify comments about his videos' background music as part of experiments on how his audience might respond differently to different genres, explaining that ``\textit{over time, I started maybe using a different kind of music. It was good to quantify a progression of comments across videos.}''
Participants also experienced joy in curating comments with \textsc{\YouTubeSystem{}}.
For instance, P6 runs a channel about her travel experiences and created filters to find and reply to sentimental comments where ``\textit{people shared their nice stories about a place where I visited.}''
P5 also mentioned feeling joy, saying, ``\textit{I think using this filter, it made me smile multiple times because I saw old comments I had forgotten about.}''

\textbf{Continued Use}. In the final interviews, we asked participants whether they would continue using \textsc{\textsc{\YouTubeSystem{}}} in their daily workflows. Out of 10 participants, 6 expressed interest in continued use.
Several found the system particularly helpful for exploring comments across videos, including older ones that they might have otherwise overlooked.
As P6 explained,``\textit{My comments go back to 2012 or somewhat...it has been a good decade. I can go back and see what people have been saying in [\textsc{\textsc{\YouTubeSystem{}}}], and it was almost a nice refresher for me of how to keep myself on track.}''
Others emphasized \textsc{\YouTubeSystem{}}'s value for managing viral videos that attracted much attention. 
As P2 noted, ``\textit{It's most helpful whenever I have a video that goes a little bit more viral and gets a lot more comments, so it's a lot easier to filter out what is worth highlighting.}''
Participants who did not plan to continue using \textsc{\YouTubeSystem{}} typically cited low comment volume as their reason. 
For example, P7 explained, ``\textit{I think if I were to get way more comments in the future, I would use the system more, but for now, I'm able to manually look at all of them. So I just think it might be like extra hassle for me to go through this program to look at the comments, versus just looking at them as they come in.}''



\subsubsection{\textbf{Does \textsc{\YouTubeSystem{}} support content creators to author interpretable and performant filters in the wild?}}\hfill

\textbf{Perceived Filter Performance}.
Participants found that they could create performant filters easily through initial bootstrapping and continual iterations in \textsc{\textsc{\YouTubeSystem{}}}, especially given its ability to accommodate their often nuanced content preferences.
For instance, P7 created two conceptually similar filters---suggestions and constructive feedback---and was impressed that \textsc{\textsc{\YouTubeSystem{}}} was able to distinguish between the two categories of comments.


However, participants noted that  \textsc{\YouTubeSystem{}} filters for preferences that require context beyond the content of comments are not very performant.
For example, P5 reported, ``\textit{I asked for comments that reference content within the video, but it seemed like the system did not understand what was in the video.}''
Similarly, P3 created a filter to remove comments irrelevant to their corresponding videos but found that it was largely ineffective.
He then suggested that, using \textsc{\YouTubeSystem{}}, ``\textit{It's a bit harder to filter comments that are irrelevant. So it's good to add the video information when filtering comments.}''
Participants also wished that \textsc{\textsc{\YouTubeSystem{}}} would incorporate more information about a comment in its decision-making process, including its number of likes and dislikes, the number of comments that its author has posted (for detecting spam), or the discussion thread in which it sits.
For instance, P4 mentioned a time when two users were arguing in a comment thread; while their insults indeed fell within the scope of personal attacks, her filter failed to recognize that these insults were not targeted at her specifically.



\textbf{Filter Interpretability}.
As illustrated in Figure \ref{usability}, participants found their filters interpretable, enabling them to better evaluate the performance of their filters and decide which ones to deploy.
For instance,  P2 found his initialized filter decently accurate after observing that the calibration process ``\textit{added on [several rubrics] to what [he] was trying to get it to catch, [such as] the political agendas and the conspiracy theory.}''
\textsc{\YouTubeSystem{}}'s filter descriptions also helped participants better understand why their filters caught or did not catch each comment.
As P4 said, ``\textit{I think the fact that [the filter description] is longer is actually helpful, because it's specific about what exactly it's catching, instead of it being really vague.}''
P7 actively reviewed filter decisions in light of their filters' descriptions: ``\textit{I remember reading the descriptions for these filters and being like, 'Oh, that's how they caught this comment.'}''
When a few filter decisions appeared mysterious, participants also appreciated that \textsc{\textsc{\YouTubeSystem{}}} generated explanations of comment classifications for their review.
Some participants suggested ways to further improve filter descriptions.
For instance, P4 complained that some ended up verbose. She proposed ``\textit{highlighting the key parts of the description}'' or summarizing each rubric into a short name.
P5 also noted that filter descriptions extensively used logical connectors like ``or,'' ``and,'' and ``not,'' suggesting that \textsc{\YouTubeSystem{}} better convey such logic with visual aids.

\subsubsection{\textbf{How do content creators find the \System{} workflow within \textsc{\YouTubeSystem{}} to initialize and iterate on their content filters?}}
Participants reported that \textsc{\textsc{\YouTubeSystem{}}} helped them easily initialize filters that captured their nuanced preferences (see Figure \ref{usability}). Many began with abstract ideas, sometimes sparked by a random comment that they encountered. They appreciated being able to provide a draft description or even a single example comment, which \textsc{\YouTubeSystem{}} then translated into a more precise filter description.
Labeling a set of 20 interesting comments also assisted participants in further mapping out vague preferences.
Recalling the labeling experience, P2 described, ``\textit{I thought I had a good idea of what kinds of comments I want to catch when I created the filter, but then when I had to make the selection of `catch' and `don't catch,' I actually realized I myself don't have very clearly formed preferences.}''
Many participants appreciated that they did not need to manually write nuances in their preferences but could simply label examples in the initialization stage.
At the same time, participants observed that some filters were common across their niche YouTube communities and suggested additional ways to initialize filters. In particular, they proposed starting from in-built templates or adapting filters shared by fellow creators.

Beyond initialization, participants liked being able to provide conceptual guidance in multiple ways when iterating on their filters. 
As P9 mentioned, ``\textit{When improving the filters, we can watch those particular comments, and clarify...if there's a second thought in my mind, or additional details I might not have exactly got in the initial filter. That feature actually helped me a lot: the more you refine, the more accurate the results are.}''
P1 also noted that iterating on his filters helped adapt to constant shifts in how his audience commented: ``\textit{I think refining that helps, because it would help the AI to then understand these new types of comments and filter those accordingly.}''
However, we also found that the degree of control participants wanted varied with the purpose and stakes of their filters. 
When filters directly affected their public channels, participants preferred finer control; when filters were used mainly for personal content curation, they favored ease of use.
As P10 explained, ``\textit{The easiest thing would be the best thing. I want to filter comments to get a better sense of some specific questions I want to ask [my audience], so that extra control isn't going to matter as much,}'' whereas ``\textit{if you're going to be removing something, you'd want it to be pretty precise.}''
Participants also sought more control when their preferences were highly nuanced or personal.
As P9 articulated, ``\textit{There are a lot of other things which are conventionally abusive words or phrases. Once we feed our labels to the machine, I feel it will make correct edits. But for the personal things, I'd sit myself and improve the filter.}''


\section{Discussion}

\subsection{User Authoring of Personal Content Filters in Real-World Usage}

Personal content classification systems arise as a critique of centralized, platform-wide moderation~\cite{jhaver2022designing}. While many of these personal systems allow users to tailor moderation to their own preferences, they also place the burden on individuals to build filters from scratch~\cite{jhaver2023personalizing}. 
Yet moderation preferences are not always unique. 
Users with shared identities or experiences--such as disabled communities or online content creators--often face similar forms of harm~\cite{heung2024vulnerable, heung2022nothing}. 
Consistent with this, we observed in our field study that participants often created similar filters to block hate speech or spam.
These similarities suggest opportunities for both research and design.
Researchers have developed hate-speech taxonomies for various user communities~\cite{heung2024vulnerable, heung2022nothing}, but there have not been many efforts to translate these taxonomies into performant filters accessible to users.
We envision that community members could also adopt or build upon filters authored by peers with similar experiences. 
In such a future, \textsc{\YouTubeSystem{}} could not only facilitate the creation of personal filters but also support the sharing and evaluation of community filters. The interpretability of its prompts could be especially valuable, enabling users to compare filters, assess their assumptions, and decide which best fit their own preferences.
It would also be particularly useful for content creators with small channels, which have fewer relevant comments than what \System{} requires to even initialize a filter with active learning. 

Our findings also highlight the complex and volatile settings of real-world content classification. Content creators increasingly maintain presences across multiple platforms such as TikTok, YouTube, and Instagram~\cite{ma2023multi}. Several participants in our study struggled not only to manage distributed comments in one place but to account for each platform’s distinct dynamics. While \textsc{\YouTubeSystem{}} could be extended as a cross-platform portal for comment classification, future research should investigate how creators might want different strategies to address varied platform contexts.
Creators in our field study also considered many contextual factors beyond the text of individual comments, as discussed in previous research~\cite{seering2023moderates}. Prior work has also explored account-level curation~\cite{geiger2016bot}, and in our field study participants additionally weighed which video a comment was posted on, how many likes and dislikes it received, and its surrounding discussion thread. 
In contrast, existing personal content filters treat each comment independently. Future work should examine how such contextual factors can be surfaced and modeled to better support user decision-making in content classification.

Finally, despite growing interest in LLMs for personal content classification, our field study highlights that deployment costs remain a major challenge.
In our field study, we estimated an average cost of \$20–30 per participant per week for LLM API usage. We implemented various measures to reduce expenses, but these constraints occasionally affected usability. For example, during onboarding we limited retrieval to the most recent 1,000 comments instead of participants’ full comment history to avoid excessive costs. Participants with over 10,000 comments noted that the system would have been more useful if it could process all of their comments. 
Similarly, we restricted the number of candidate prompt edits explored in the optimization workflow, since evaluating each candidate required as many LLM requests as the total number of user labels in our classifier implementation.
While our research lab was able to support these costs for the duration of the study, sustaining such expenses at scale would be prohibitive. 
Future research should explore techniques such as LLM cascading or model distillation to maintain performance while reducing costs. For instance, cheaper models could handle less complex or lower-stakes preferences~\cite{chen2023frugalgpt}, while distilled lightweight classifiers could support heavy users without requiring repeated LLM calls~\cite{xu2024survey}.

\subsection{Implications of Human-in-the-Loop Prompt Optimization Beyond Content Classification}

Research on human–AI collaboration has long debated whether combining human guidance and AI assistance leads to performance better than either can attain alone~\cite{lai2022human,hemmer2025complementarity}. Some studies also warn that human involvement can introduce biases~\cite{zipperling2025s}, over-reliance~\cite{bansal2021does}, or shallow engagement with model outputs~\cite{poursabzi2021manipulating}.
Our results echo this broader tension: we did not observe significant performance gains from user intervention 
compared to automatic prompt optimization.
This may reflect limitations in our experiment design. As participants were only asked to imagine that they were managing YouTube comments, they might not always have had clear, targeted feedback during iterations. Several defaulted to addressing the first mistake that they saw, rather than systematically refining their filters.
It might also be possible that we can observe more performance gains after several rounds of iterations, or when there is a explicit distribution shift of the comments~\cite{li2023robust, lai2022human}.

Despite no changes in measured performance, users still desired the control and interpretability provided and even perceived that performance was better in \System{}.
Our results also point toward additional opportunities to design richer forms of human–AI collaboration in prompt optimization. 
Rather than only clustering mistakes, future systems could generate a semantic graph of errors, enabling users to see connections and identify emerging failure patterns. 
Our current workflow also assumes a one-way interaction: users can review and select iterations, but cannot critique or refine them further. Likewise, the underlying models lack awareness of prior failed attempts. 
Designing interactions where users and models build on one another’s feedback may offer a path toward higher-quality prompt optimizations.

Additionally, the \System{} workflow could potentially generalize beyond content classification to other LLM tasks, especially more open-ended ones where user guidance is crucial~\cite{feng2024policy}. 
For example, users naturally audit their personal writing assistants or chatbots over time~\cite{guo2025pen, wang2025social}, and this accumulated feedback could similarly be aggregated into failure patterns for review and prompt optimization.
However, open-ended tasks such as writing assistance or chatbots raise new challenges for evaluating prompt quality. Whereas the performance of content classifiers can be measured against a ground-truth dataset, users in open-ended tasks often rely on more ambiguous criteria and struggle to provide ``golden'' examples~\cite{clark2021all, gehrmann2023repairing}. This makes automatic evaluation of prompt iterations far more difficult. One emerging direction is to leverage LLMs themselves as judges, applying user-defined criteria to evaluate candidate prompts~\cite{kim2024evallm, shankar2024validates}.
Looking ahead, prompt optimization need not be limited to a single user. We envision settings where groups of people collectively deliberate over guidance and review generated candidates. Such deliberation can support consensus-building and create a more interactive optimization process, which may be especially valuable for contexts like community chatbots or the development of LLM policies~\cite{feng2024policy, sharma2024experts}.

\section{Limitations}

Our experiments have some limitations.
First, we restricted our test dataset to 100 comments to keep our experiments a reasonable length, which may have limited our evaluation's accuracy.
Second, our lab sample consisted primarily of undergraduate and graduate students, who are not representative of the broader population of social media users. Future work should conduct larger-scale studies with more diverse participants to validate and extend our findings.
We also constrained the computational resources allocated to optimization algorithms in order to support real-time interactions. Future work could explore whether more powerful optimization methods would better support user iteration or even reduce the need for manual intervention.

Our field study likewise has limitations. While we focused on YouTube creators, other groups may benefit from our workflow but experience it differently. For example, casual social media users may find the workflow too demanding compared to content creators, who are more motivated to manage comments~\cite{jhaver2022designing}.
Future research should explore integrating our workflow into everyday ``teachable feed'' experiences that allow users to train classifiers through lightweight, ongoing engagement~\cite{feng2024mapping}.
Our sample of 10 participants was small, and content creators themselves are a diverse group with specialized needs~\cite{heung2025ignorance}; studying how particular creator communities interact with \textsc{\YouTubeSystem{}} could inform tailored design interventions.
Finally, our field study lasted three weeks due to resource constraints and thus did not observe the long-term utility of \textsc{\YouTubeSystem{}}, such as how filters evolve across a wider range of comments and shifting community norms.
A future longitudinal study could better capture these dynamics over extended use.



\section{Conclusion}

In this paper, we introduced \System{}, a human-in-the-loop prompt optimization workflow built on the key insight that user steering and prompt optimization should be decoupled. During initialization, users can label examples to express their intuitive preferences; over time, they can provide high-level conceptual feedback to iteratively refine their filters. 
At each stage, \System{} translates this input into candidate prompts that are performant, generalizable, and interpretable for user review. 
In a controlled lab experiment, participants unanimously preferred \System{} over a state-of-the-art automatic prompt optimization baseline, citing its support for diverse iteration strategies and significantly greater interpretability. 
A three-week field study with YouTube creators further showed how \YouTubeSystem{} enabled the creation of diverse, high-performing filters that supported both audience engagement and community protection. 
Looking ahead, we see opportunities to extend our workflow to more open-ended LLM tasks and collective settings, highlighting the importance of interpretability and user-led steering in designing sustainable and trustworthy LLM-powered applications.

\begin{acks}
This research was supported by NSF award \#2236618. We would like to thank members of the Social Futures Lab at the University of Washington for their invaluable help in this project. We also would like to thank our anonymous reviewers for their insightful feedback. Finally, we would like to express our heartfelt thanks to all the participants who dedicated their time and effort to participate in our study.
\end{acks}


\bibliographystyle{ACM-Reference-Format}
\bibliography{filterbuddy}

\appendix

\section{Difficulties of Recruiting Content Creators}\label{discussion:recruitment}
Our field study highlights a critical challenge for designing and evaluating tools like \textsc{\YouTubeSystem{}}: recruiting content creators. Finding and building trust with this community is difficult, and these challenges have implications for future research in this area.

One major hurdle is that there are no clear best practices for recruiting content creators.
In our field study, we relied on a variety of methods used in prior work, including directly messaging creators~\cite{steen2023you, jhaver2022designing} and posting in online groups~\cite{ma2023multi}. Both were time-consuming and inefficient. 
We emailed over 200 creators, with only 5 ultimately enrolling in the study. 
In one case, we were banned from a Discord server for mistakenly posting our recruitment message without moderator approval, violating community guidelines. 
This emphasizes the importance of understanding and respecting the norms of a community before trying to recruit from it.
Even with these understandings, there might still be little success. In total, we contacted moderators of 17 Discord servers, but only 3 granted permission to post our recruitment message.

Another difficulty was building trust. Many online community members were suspicious of our team, mistaking us for bots or spammers, even after we sent links to our academic profiles to prove our credibility. 
In one Discord community, moderators gave us a
``researcher'' badge to vouch for us, but some members still did not trust us, even though people typically become more receptive to a researcher when the latter has been vouched for by a trusted connection as trustworthy~\cite{small2009many}. 
Other members simply found our presence odd, which expands prior research on being aware of a community's culture for optimal and respectful recruitment~\cite{goodwin2023using}.
We found that casually chatting with channel members for a few weeks helped build rapport, but this is not a feasible strategy for time-constrained studies.

Reciprocal trust was also an issue, as we struggled to verify that people were not bots, spammers, or people lying to get compensation. 
Although we used a pre-screening survey, about 90\% of responses were nonsensical, indicating that they were likely spam. 
We believe that some respondents may have used LLMs to craft convincing responses, a problem we only caught when they could not log into their YouTube accounts during onboarding. 
If we had not caught this with an on-call credibility check, it would have undermined the credibility of our findings about the utility of \textsc{\YouTubeSystem{}} for content creators.
Future research could explore ways to use LLMs themselves to detect such deception, for example by scanning pre-screening survey responses for indicators of non-human deception like pathological positivity~\cite{huang2025decoding}.
Additionally, systems that facilitate meronymous communication, where people systematically reveal conscientiously chosen details or rely on trusted endorsers to vouch for their credibility~\cite{soliman2024mitigating}, could foster reciprocal trust between researchers and potential participants when recruiting online for system design and usability studies.


\end{document}